\documentclass[sigconf]{acmart}

\usepackage{subcaption}
\usepackage{multirow}
\usepackage{bm}
\usepackage{bbding}

\copyrightyear{2023}
\acmYear{2023}
\setcopyright{acmlicensed}\acmConference[SIGIR '23]{Proceedings of the 46th
International ACM SIGIR Conference on Research and Development in
Information Retrieval}{July 23--27, 2023}{Taipei, Taiwan}
\acmBooktitle{Proceedings of the 46th International ACM SIGIR Conference on
Research and Development in Information Retrieval (SIGIR '23), July 23--27,
2023, Taipei, Taiwan}
\acmPrice{15.00}
\acmDOI{10.1145/3539618.3591681}
\acmISBN{978-1-4503-9408-6/23/07}

\begin{document}


\title{EulerNet: Adaptive Feature Interaction Learning via \\ Euler's Formula for CTR Prediction}

\author{Zhen Tian$^{\dagger}$}
\affiliation{%
  \institution{Gaoling School of Artificial Intelligence, Renmin University of China}
  \city{Beijing}
  \country{China}
}
\email{chenyuwuxinn@gmail.com}

\author{Ting Bai$^{*\ddagger}$}
\affiliation{%
  \institution{Beijing University of Posts and Telecommunications}
  \city{Beijing}
  \country{China}
}
\email{baiting@bupt.edu.cn}

\author{Wayne Xin Zhao$^{*\dagger}$}
\affiliation{%
  \institution{Gaoling School of Artificial Intelligence, Renmin University of China}
  \city{Beijing}
  \country{China}
}
\email{batmanfly@gmail.com}

\author{Ji-Rong Wen$^\dagger$}
\affiliation{%
  \institution{Gaoling School of Artificial Intelligence, Renmin University of China}
  \city{Beijing}
  \country{China}
}
\email{jrwen@ruc.edu.cn}

\author{Zhao Cao}
\affiliation{%
  \institution{Poisson Lab, Huawei}
  \city{Beijing}
  \country{China}
}
\email{caozhao1@huawei.com}

\thanks{$^*$ Ting Bai (baiting@bupt.edu.cn) and Wayne Xin Zhao (batmanfly@gmail.com) are the corresponding authors.}
\thanks{$^\dagger$ Also with Beijing Key Laboratory of Big Data Management and
Analysis Methods}
\thanks{$^\ddagger$ Also with Beijing Key Laboratory of Intelligent Telecommunications Software and Multimedia}

%

\newcommand{\ie}{\emph{i.e.,} }
\newcommand{\eg}{\emph{e.g.,} }
\newcommand{\paratitle}[1]{\vspace{1.5ex}\noindent\textbf{#1}}

\newcommand{\modified}[1]{\textcolor{blue}{#1}}

\renewcommand{\authors}{Zhen Tian, Ting Bai, Wayne Xin Zhao, Ji-Rong Wen and Zhao Cao}
\renewcommand{\shortauthors}{Zhen Tian et al.}

\begin{abstract}
Learning effective high-order feature interactions is very crucial in the CTR prediction task.
However, it is very time-consuming to calculate high-order feature interactions with massive features in online e-commerce platforms.
Most existing methods manually design a  maximal order and further filter out the useless interactions from them.
Although they reduce the high computational costs caused by the exponential growth of high-order feature combinations, they still suffer from the degradation of model capability due to the suboptimal learning of the restricted feature orders.
The solution to maintain the model capability and meanwhile keep it efficient is a technical challenge,  which has not been adequately addressed.
To address this issue, we propose an adaptive feature interaction learning model, named as \textbf{EulerNet}, in which the feature interactions are learned in a complex vector space by conducting space mapping according to Euler's formula.
EulerNet converts the exponential powers of feature interactions into simple linear combinations of the modulus and phase of the complex features, making it possible to adaptively learn the high-order feature interactions in an efficient way.
Furthermore, EulerNet incorporates the implicit and explicit feature interactions into a unified architecture, which achieves the mutual enhancement and largely boosts the model capabilities.
Such a network can be fully learned from data, with no need of pre-designed form or order for feature interactions. 
Extensive experiments conducted on three public datasets have demonstrated the effectiveness and efficiency of our approach.
Our code is available at: \textcolor{blue}{\url{https://github.com/RUCAIBox/EulerNet}}.
\end{abstract}



\maketitle

\section{Introduction}
Click-Through Rate (CTR) prediction, which aims to predict the probability of a user clicking on an item, is a very critical task in  online e-commerce platforms. 
In the literature, various approaches have been proposed for effective CTR prediction~\cite{cheng2016wide,he2017neural,li2020interpretable,shan2016deep}. 
The key of CTR prediction is to accurately model the complicated context data by capturing  underlying feature relationships.  
Typically, these methods either learn \emph{explicit feature interaction} by manually setting  the interaction form/order via factorization based models~\cite{pan2018field, sun2021fm2}, or \emph{implicit feature interaction} by directly 
modeling the fusion of all the features via deep neural networks~\cite{cheng2016wide, zhang2016deep}. 


Despite the progress, these methods still have limitations in learning complicated feature relationships (\eg high-dimensional varied contexts).  Firstly, due to an exponential growth of combinational complexity,  explicit learning methods usually set a small interaction order, which cannot scale to the cases requiring high-order feature interaction modeling. Further, they only model the integer-order interactions, thus leading  to an inaccurate modeling of real-world scenarios.   Secondly, due to the lack of effective design in interaction mechanisms, implicit learning methods are shown to be less effective than explicit learning methods~\cite{rendle2020neural}.
  
A major challenge in modeling high-order interactions among raw features is the incurred high computational cost due to the exponential feature combinations as the number of raw features increases.
In a practical scenario, raw features tend to be very sparse and have hundreds of fields with millions of dimensions.
For example, identifier features like user ID  or item ID become very sparse when encoded as one-hot vectors, so are the multi-field features extracted from the user behavior logs.
Calculating high-order interactions on such sparse features with hundreds of fields is computationally intensive and time-consuming.

Considering the above limitations, several studies~\cite{lian2018xdeepfm, tian2023directed, wang2021dcn}  manually assign a  maximal order, and further remove useless  interactions from them. 
However, they still suffer from the degradation of model capability due to the restricted feature orders. 
As a promising approach, a recent study AFN~\cite{cheng2020adaptive} leverages logarithmic neural network (LNN~\cite{hines1996logarithmic}) to adaptively learn the order of feature interactions. It can automatically learn the orders of feature interactions, but at the expense of limited feature representation space, \ie only positive feature embeddings can be learned in logarithmic space transformation, which requires a large consumption of logarithmic neurons for retaining the performance.

To address these issues, in this paper, we  propose an adaptive feature interaction learning model, named as \textbf{EulerNet}, for automatically learning arbitrary-order feature interactions. 
Unlike prior work, the core idea of EulerNet is to model the feature interaction in a \emph{complex vector space} by conducting  space mapping according to Euler’s formula. 
Specially, EulerNet converts the exponential powers of feature interactions into simple linear combinations of the modulus and phase of the complex features, making it feasible to capture complicated feature interactions in an \emph{efficient, flexible} way. 
Based on such an idea, we develop an Euler interaction layer that performs the above transformation, which can be stacked to form a capable interaction learning network.
Such a network can be fully learned from data, with no need of pre-designed form or order for feature interactions. 
Furthermore, Euler interaction layer can be extended to integrate the implicit feature interactions. 
Different from previous  explicit-implicit hybrid approaches, our model can  fuse the feature representations from the two ways in the Euler interaction layer, instead of simply keeping  two separate feature interaction models.   

The contributions are summarized as follows:




$\bullet$ We propose an adaptive feature interaction learning model EulerNet. It can automatically learn the arbitrary-order feature interactions from data.  Meanwhile,  our model can  jointly capture the explicit and implicit feature interactions in a unified model architecture. 


$\bullet$ We propose to model the feature interaction in the complex vector space, by conducting space mapping  according to Euler's formula.  It enables EulerNet to convert the complicated 
exponential powers into simple linear computation.

$\bullet$ We conduct extensive experiments on three widely used datasets. EulerNet consistently outperforms a number of competitive baselines with much fewer parameters, showing the effectiveness and efficiency of our model.

\section{Preliminary} \label{sec:exh}
We first introduce the  CTR prediction task,  then present the formulations for  explicit and implicit feature interactions in existing work, and finally introduce the Euler's formula used in our model.


\paratitle{CTR Prediction.}
The task of the click-through rate~(CTR) prediction aims to estimate the probability that 
a user will click on an item. It takes as input a vector of context features (\eg user and item features), denoted as $\bm x = \{ x_1,  x_2,..., x_m \}$, where  $m$ is the number of feature fields and $x_j$ is the $j$-th feature, the label $y \in \{0 , 1\}$  represents whether the item is clicked or not and it is predicted from the input feature $\bm x$. 
We further apply a look-up operation to each feature ${x}_j$ by mapping it  
into a $d$-dimensional embedding $\bm{e}_j \in \mathbb{R}^d$.  
In this way, the original feature vector can be represented as a list of feature embeddings $\{ \bm{e}_1,\bm{e}_2,..,\bm{e}_m \}$. 

\paratitle{Explicit Feature Interactions.}\label{sec:expf}
The key of CTR prediction is to learn the effective feature interactions, which is a fundamental problem for this task~\cite{zhang2021deep}. 
According to the interaction forms, existing methods can be roughly divided into \emph{explicit} and \emph{implicit} feature interactions.  
Explicit feature interactions is usually modeled by a pre-designed interaction formula with a controllable order, such as FM~\cite{rendle2010factorization}, HOFM~\cite{blondel2016higher} and IM~\cite{yu2020deep}.
We introduce a special symbol $\Delta_{ex}$ to denote the explicit feature interaction,  generally defined as: 
\begin{equation}\label{eq-ex-int}
\Delta_{ex}=
 \sum_{ \alpha \in \mathcal{A}} \bm e_1^{\alpha_1} \odot \bm e_2^{\alpha_2} \odot \cdots \odot \bm e_m^{\alpha_m},
\end{equation}
where $\bm \alpha = [\alpha_1, \alpha_2, ..., \alpha_m]$ consists of the orders for each feature in $\bm{x}$,  $\odot$ is the element-wise product. 
Based on  $\Delta_{ex}$, another prediction  function $f(\cdot)$ (\ie sigmoid function) can be employed to generate the predicted label $\hat{y}$ in $[0,1]$. Here, $\mathcal{A}$ is the  set of all planned interactions by a CTR model.  
Most CTR models require the interaction orders to be non-negative integers,  \ie  $\alpha_j  \in \mathbb{N}^0$. 
 For example, FM~\cite{rendle2010factorization} only considers second-order interaction, which specify $\mathcal{A} = \{ \bm{\alpha} | \sum_{j=1}^{m} \alpha_j = 2, \forall \alpha_j \in \{0, 1\}\}$.  
Different from most existing methods~\cite{sun2021fm2, rendle2010factorization, pan2018field}, we aim to learn the arbitrary-order feature interactions in an adaptive learning way, \ie $\bm \alpha$ could be arbitrary real values that are automatically learned from data.

\paratitle{Implicit Feature Interactions.}\label{sec:impf}
As another form of feature interaction, implicit feature interactions are commonly modeled by feed-forward neural networks, \eg the multi-layer perceptron (MLP) used in xDeepFM~\cite{lian2018xdeepfm}, DCNV2~\cite{wang2021dcn} and DeepIM~\cite{yu2020deep}. 
Different from explicit feature interactions, it does  not specify the concrete interaction forms  in the model.   Formally, given the concatenation of all feature embeddings, \ie ${\bm z}^{(0)} = [\bm e_1; \bm e_2; ... ; \bm e_m]$,
the implicit feature interaction process $ \Delta_{im} $ can be formulated as:
\begin{align}
\Delta_{im} &=  \bm{z}^{(L)}, \\
  \bm{z}^{(l)} &= \sigma(\bm W^{(l)}  \bm{z}^{(l-1)} + \bm b^{(l)} ),
\end{align}
where $l \in [1, L]$, $L$ is the layer depth and $\sigma$ is the activation function.

\paratitle{Euler's Formula.}\label{sec:impf}
Euler's formula is a mathematical formula that establishes the relationships between different expressions of complex vectors, and can be formulated as:
\begin{align}
  \bm \lambda e^{i \bm \theta} = \bm \lambda \cos{\bm \theta} + i (\bm \lambda \sin{\bm \theta}) \label{eq:euller},
\end{align}
where $\bm \lambda e^{i \bm \theta}$ and $\bm \lambda \cos{\bm \theta} + i (\bm \lambda \sin{\bm \theta})$ are the representations of a complex vector in the polar form and the rectangular form respectively. 
Here, $i$ is the imaginary unit, $\bm \lambda$ and $\bm \theta$ are the modulus and phase of a complex vector. 
For a complex vector $\bm r + i \bm p$, we set the real part $\bm r = \bm \lambda \cos{\bm \theta}$ and imaginary part $\bm p = \bm \lambda \sin{\bm \theta}$. 
The modulus $\bm \lambda$ and phase $\bm \theta$ can be represented as:
\begin{align}
\begin{split}
    \bm \lambda &= \sqrt{\bm r^2 + \bm p^2},\\
    \bm \theta &= \mathrm{atan2}(\bm p, \bm r),
\end{split}
\end{align}
where $\mathrm{atan2}(\bm y, \bm x)$ is the two-argument arctangent function. 
The transformation via Euler's formula makes it feasible to convert the complex vectors from the rectangular form to the polar form, providing a way to encode the features in the polar space.

\section{Methodology}
To adaptively learn the arbitrary-order feature interactions, we propose a feature interaction learning model via Euler's formula, named as \textbf{EulerNet}.
We first present a general introduction of our model, and then introduce the technical details in each part. 


\begin{figure}[h]
    \centering
    \includegraphics[width=1.\linewidth]{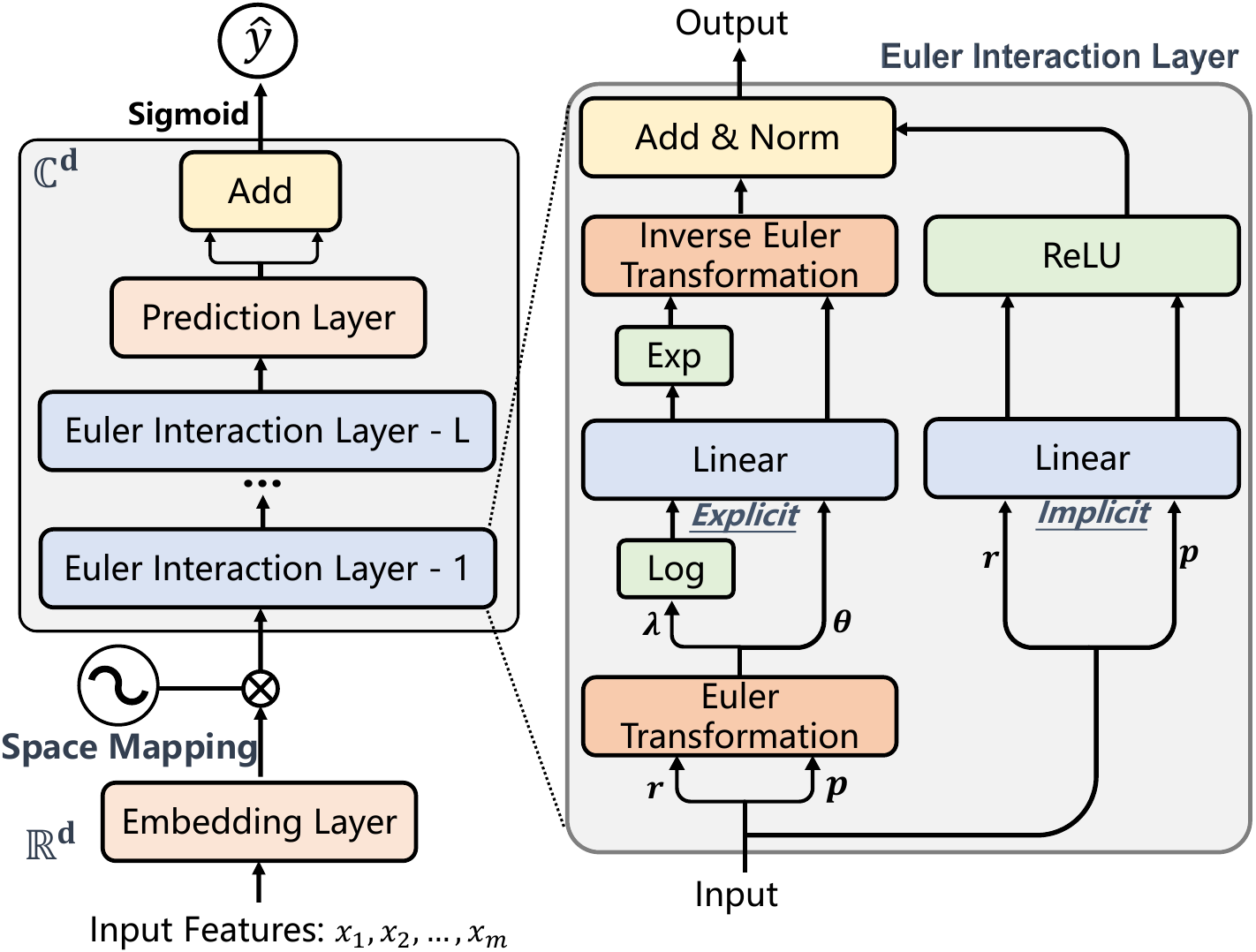}
    \caption{The overall architecture of EulerNet.}
    \label{fig:framework}
  \end{figure}

\subsection{Overview of EulerNet}
The overview architecture of EulerNet is shown in Figure~\ref{fig:framework}. 
EulerNet is designed by stacking the key structure of \emph{Euler interaction layer}. 
The core idea of Euler interaction layer is to transform explicit interaction of feature embeddings (Eq.~\eqref{eq-ex-int})  in a \emph{complex vector space} according to Euler’s formula (Eq.~\eqref{eq:euller}). As such, we can model complicated feature relationships in a flexible way, without the constraints in existing work (\eg non-negativity or integer). Further, exponential computation can be simplified as linear computation, making  it possible to adaptively learn the high-order feature interactions in an efficient way. 
Further, Euler interaction layers can be extended to incorporate implicit feature interaction learning, which can naturally integrate the two kinds of feature interaction. 

In what follows, we introduce the details of explicit feature interaction (Section~\ref{sec:explicit})
 and implicit feature interaction (Section~\ref{sec:implicit}).


\subsection{Explicit Feature Interaction Learning} \label{sec:explicit}
Previous works~\cite{rendle2010factorization, guo2017deepfm, wang2021dcn, lian2018xdeepfm} mainly learn the feature interactions in the real vector space, which 
limits the expressiveness of features, lacking  the ability to adaptively capture the arbitrary-order feature interactions.
To address this issue, we first map the input features from the real vector space to the complex vector space, and then learn the explicit feature interactions in the complex vector space. 

\subsubsection{Complex Vector Representation of Features}

As discussed in Section~\ref{sec:exh},  the original input vector $\bm{x}$ can be mapped into  a list of feature embeddings $\{ \bm{e}_1,\bm{e}_2,..,\bm{e}_m \}$ via an embedding layer. 
Based on the embedding representations, we next discuss how to map them into  complex space and further conduct Euler interaction.

\paratitle{Complex Space Mapping.}
To improve the expressiveness of features, we map the feature embeddings from the \emph{real vector space} to the \emph{complex vector space}. Given a feature embedding, 
$\bm{e}_j$, in the complex vector space, we utilize two real vectors $\bm r$ and $\bm p$ to  represent the real and imaginary parts of the complex vector respectively (\ie $\tilde{\bm{e}}_j = \bm{r}_j + i \bm{p}_j$). To transform a feature embedding  into a complex vector, the key idea is to consider it as the phase and incorporate a learnable parameter (or parameter vector) $\mu_j$ as the modulus following Euler's formula in Eq.~\eqref{eq:euller}:
\begin{eqnarray}\label{eq-transform-1}
\bm{\lambda} \rightarrow  \mu_j, \text{~~~~~~~}  \bm{\theta} \rightarrow \bm{e}_j. 
\end{eqnarray}
According to  Eq.~\eqref{eq:euller}, we can obtain the corresponding complex representation of $\bm{e}_j$ by introducing the modulus parameter $\mu_j$:
\begin{eqnarray}
\tilde{\bm e}_j=\underbrace{\mu_j \cos({\bm e}_j)}_{\text{real}} + i \underbrace{\mu_j \sin({\bm e}_j)}_{\text{imaginary}},
\end{eqnarray}
where we have $\bm{r}_j = \mu_j \cos({\bm e}_j)$ and $\bm{p}_j = \mu_j \sin({\bm e}_j)$. 
To enhance the field-specific semantics, we let the feature embeddings corresponding to the same field share the same modulus parameter. 
%
After complex space mapping, each feature is represented by a complex vector $\tilde{\bm e}_j$.
We utilize the complex feature representations $\{\tilde{\bm e}_j\}_{j = 1}^m = \{\bm r_j + i \bm p_j\}_{j=1}^m$ for subsequent interaction modeling. 


\subsubsection{Euler Interaction Layer}\label{sec:eulerint}
Euler interaction layer is the core component of our proposed EulerNet, which enables the adaptive learning of  explicit feature interactions. 
An Euler interaction layer performs the feature interaction under the complex space one time, taking as input a complex representation and outputting a transformed  complex representation. In this way, we can stack multiple Euler interaction layers for enhancing the model capacity. Next, we describe the transformation process with an Euler interaction layer. 



\paratitle{Euler Transformation.}
In order to adaptively learn the explicit feature interactions, we  utilize Euler Transformation to transform the complex feature representations from the rectangular form to the polar form. 
This step can convert exponential  multiplications into simplified linear computation, making it feasible to adaptive capture complicated feature interactions. 
Given the input complex representation $\bm r_j + i \bm p_j$ of feature embedding ${\bm e}_j$, we use Euler's formula in Eq.~\eqref{eq:euller} to obtain the polar-form representations: 
\begin{equation}
\bm r_j + i \bm p_j \rightarrow \bm{\lambda}_j e^{i \bm{\theta}_j}.
\end{equation}  
In this form, the explicit feature interaction can be formulated as: 
\begin{align} 
\begin{split}
  \Delta_{ex} &=\tilde{\bm e}_1^{\alpha_1} \odot \tilde{\bm e}_2^{\alpha_2} \odot .. \odot \tilde{\bm e}_m^{\alpha_m}\\
  &= \prod_{j = 1}^{m} \bigg(\bm \lambda_j^{\alpha_j} \exp{(i\alpha_j\bm \theta_j)}\bigg) \\
  &=  \exp\bigg( \sum_{j = 1}^{m} \alpha_j \log(\bm \lambda_j)\bigg)\exp{(i \sum_{j = 1}^{m} \alpha_j\bm \theta_j)},
\end{split}
\label{eq:log}
\end{align}
where $\bm \lambda_{j}= \sqrt{\bm r_j^2 + \bm p_j^2} $ (always non-negative) and  $\bm \theta_j= \mathrm{atan2}(\bm p_j, \bm r_j)$ are the modulus and phase vectors of the complex features in the polar form.
In this way, explicit feature interaction has been cast into a linear weighted  combination of modulus and phase values in the polar space, and the original interaction order  (\ie $\bm \alpha$) becomes the combination coefficients. 

Note that, to achieve the similar formulation,  we can also perform the log operation on the original feature interaction (Eq.~\eqref{eq-ex-int}), while it requires the feature embeddings to be \emph{non-negative}, which do not always hold for all the cases.   
Such a transform provides a possibility to model complicated feature interaction in a more simplified way.  


\paratitle{Generalized Multi-order Transformation.}
In the above, we have discussed the case with an order vector $\bm{\alpha}$. In this part, we generalize such a transformation into a group of $n$ order vectors $\{\bm  \alpha_k \}_{k=1}^n$, where $\alpha_{k,j}$ denotes the $j$-th order of the $k$-th vector $\bm \alpha_k$.  
Formally, we introduce the $\bm \psi_k$ and $ \bm l_k$ to generalize Eq.~\eqref{eq:log} as follows:
\begin{align}
\begin{split}
     \bm \psi_k &= \sum_{j = 1}^{m} \alpha_{k,j} \bm \theta_j +\bm \delta_k ,\\
 \bm l_k &= \exp \bigg(\sum_{j = 1}^{m} \alpha_{k,j} \log(\bm \lambda_j) + \bm \delta'_k \bigg),
\end{split}
\label{eq:psi}
\end{align}
where $\bm \delta_k$ and $\bm \delta'_k$ are learnable bias vectors that are incorporated for enhancing the representations. 
With this generalized extension, we can obtain the explicit  interaction  with $\bm \alpha_k$ in the polar form: 
\begin{align}
\textstyle
\begin{split}
         \Delta_{ex}       =&  \exp\bigg( \sum_{j = 1}^{m} \alpha_{k,j} \log(\bm \lambda_j) +  \bm \delta'_k\bigg) \exp{\bigg(i (\sum_{j = 1}^{m} \alpha_{k,j}\bm \theta_j+ \bm \delta_k) \bigg)}\\
                =& \bm l_k e^{i \bm \psi_k}.
\end{split}
\label{eq:single}
\end{align}

\paratitle{Inverse Euler Transformation.}
Since the above feature interactions are in the polar form, we do not directly perform the corresponding interactions with a group of multi-order coefficients $\{\bm \alpha_k \}_{k=1}^n$.  
We further utilize \emph{inverse Euler transformation} to convert them into the original complex vectors in the rectangular form as:
\begin{align}
\begin{split}
\hat{\bm r}_k &= \bm l_k \cos(\bm \psi_k)\label{eq:ri},\\
\hat{\bm p}_k &= \bm l_k \sin(\bm \psi_k),
\end{split}
\end{align}
where $\hat{\bm r}_k$ and $\hat{\bm p}_k$ are the real and imaginary vectors.
In this way, the Euler interaction layer can model $n$ explicit feature interactions.
The generalized explicit feature interactions with a group of multi-order coefficients $\{\bm \alpha_k \}_{k=1}^n$ learned in Euler interaction layer can be described as:
\begin{align}
\begin{split}
     \Delta_{ex}
  &= \sum_{k = 1}^n \bm l_k \cos(\bm \psi_k) + i (\bm l_k \sin(\bm \psi_k))\label{eq:group}\\
  &= \sum_{k = 1}^n (\hat{\bm r}_k + i \hat{\bm p}_k).
\end{split}
\end{align}

This formula is the core of the  proposed EulerNet model for explicit feature interactions. Unlike prior work, the order of the interactions (\ie $\alpha_{k,j}$) can be set to arbitrary real value, without additional limits such as \emph{non-negativity}. Instead of  manually setting the order coefficients, we adaptively learn them from data, and  use the number of  order vectors $n$ to  control  the model complexity. Furthermore, we can also set varying $n$ at  different layers to increase the model flexibility.  





\subsection{Integrating Implicit Interactions}\label{sec:implicit}
Considering that  feature relationship in the real scenarios is very complicated, we further incorporate implicit feature interactions into our model.
Different from previous studies~\cite{guo2017deepfm, wang2021dcn, lian2018xdeepfm}, which model the explicit and implicit feature interactions in different architectures, we integrate them in each Euler interaction layer to enhanced the representation capacity.

\subsubsection{Fusing Explicit and Implicit Interactions}
To model more  complicated feature relationship, we construct a neural network component for capturing implicit feature interactions. 
Given the input complex features $\{\tilde{\bm e}_j\}_{j=1}^m = \{\bm r_j + i \bm p_j\}_{j=1}^m$, we can obtain the input of the implicit interaction  by concatenating the these vectors as:    $\bm r= [\bm r_1; \bm r_2, ...; \bm r_{m}]$ (\emph{real part}) and $\bm p = [\bm p_1; \bm p_2, ...; \bm p_{m}]$ (\emph{imaginary part}).
Then,  we feed the real and imaginary parts of feature representations  into the same linear layer with a subsequent non-linear activation function: 
\begin{align}
\begin{split}
    \bm r_k' &= \mathrm{ReLU}(\bm W_{k} \bm r + \bm b_k),\\
\bm p_k' &= \mathrm{ReLU}(\bm W_{k} \bm p + \bm b_k),
\end{split}
\label{eq:impu}
\end{align}
where $k \in \{1,\cdots n\}$, and $\bm W_k \in \mathbb{R}^{d \times md}$ is the weight matrix and $\bm{b}_k \in \mathbb{R}^d$ is the bias.
Finally, in order to integrate the two kinds of feature interaction, we add the explicit and implicit representations (See Eq.~\eqref{eq:ri} and Eq.~\eqref{eq:impu}) by real and imaginary parts accordingly as: 
\begin{align}
\{{\bm o}_k\}_{k = 1}^{n} = \{(\hat{\bm r}_k + \bm r_k') + i (\hat{\bm p}_k + \bm p_k')\}_{k = 1}^{n}.
\end{align}

We can stack multiple Euler interaction layers by taking the output features of the previous layer as the input for the next layer and optionally applying normalization methods such as BatchNorm~\cite{ioffe2015batch} or LayerNorm~\cite{ba2016layer} to adjust the distribution as it passes through each layer.


\subsubsection{Output for CTR Predictions}\label{sec:output}
In order to predict the CTR value, we further perform linear regression on the output representations  $\tilde{\bm o} = \{ \tilde{\bm o}_k\}_{k = 1}^{n} = \{\tilde{\bm r}_k + i \tilde{\bm p}_k\}_{k = 1}^{n}$. 
Specially, we concatenate  the real and imaginary vectors accordingly, and introduce a regression weight vector $\bm w \in \mathbb{R}^{nd}$, so as to obtain a scalar value for both the real and imaginary parts: 
\begin{equation}
z =  \bm w^{\top} \tilde{\bm r} + i (\bm w^{\top} \tilde{\bm p}) = z_{re} + i  z_{im},
\end{equation}
where $z_{re}$ and $z_{im}$ are the real and imaginary part of $z$ respectively. 
The prediction for CTR by integrating both explicit and implicit interactions can be given as:
\begin{equation}
  \hat{y} = \sigma(z_{re} + z_{im}).
\end{equation}

For training, we utilize the binary cross entropy loss with a regularization term to train our model, which is formulated as:
  \begin{equation}
    \mathcal{L}(\Theta) = -\frac{1}{N} \sum_{j=1}^{N}\bigg(y_i\log(\hat{y}_j)+(1-y_j)\log(1-\hat{y}_j)\bigg) + \gamma||\Theta||^2_2,
  \end{equation}
where $y_j$ and $\hat{y}_j$ are the ground-truth label and predicted result of $j$-th training sample respectively, and $\Theta$ denotes the set of the parameters and $\gamma$ is the $L_2$-norm penalty.

\subsection{Discussion} \label{sec:approach}
\subsubsection{Intuitive Explanation of Feature Interaction}\label{sec:enhanceana}
To have an intuitive understanding of our approach, we consider a simple case when the embedding dimension $d=1$.  Further, since we apply normalization at the Euler interaction layer, the modulus is around 1, so that we can omit the corresponding $\lambda$ from Eq.~\eqref{eq:log}. The forms of explicit and implicit interaction can be  simplified as:
\begin{align}
     g_{ex}(\tilde{\bm e}_j, \tilde{\bm e}_k)&= \tilde{\bm e}_j^{\alpha_j} \odot \tilde{\bm e}_k^{\alpha_k}  \approx \exp{\big(i(\alpha_j\bm \theta_j + \alpha_k\bm \theta_k)}\big),\\
 g_{im}(\tilde{\bm e}_j, \tilde{\bm e}_k)&=\mathrm{ReLU}(W_j\bm r_j + W_k\bm r_k) + i \mathrm{ReLU}(W_j\bm p_j + W_k\bm p_k).
\end{align}

\begin{figure}[!h]
  \centering
  \includegraphics[width=1.\linewidth]{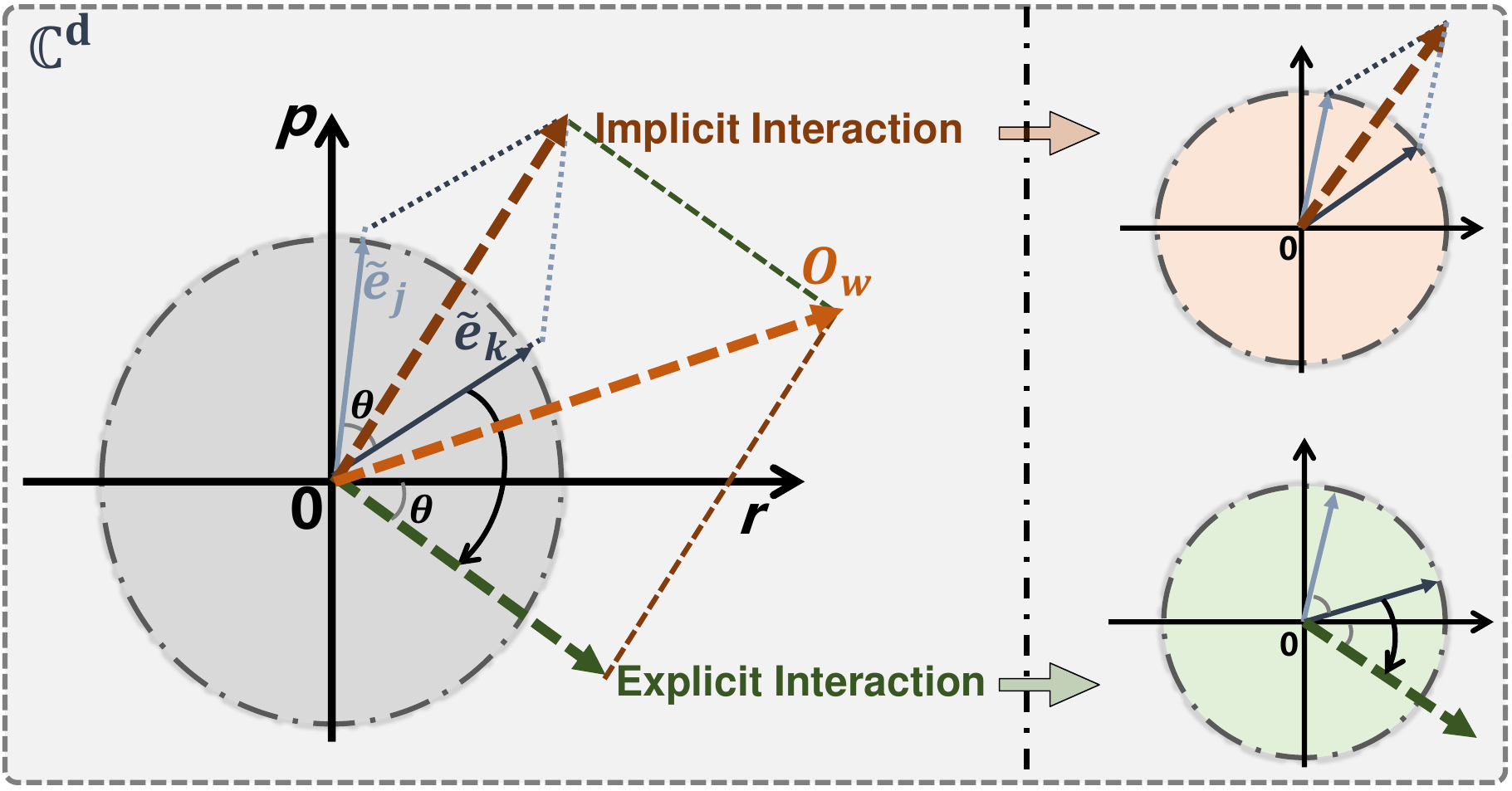}
  \captionsetup{font={small}}
  \caption{The visual understanding of the interactions in EulerNet.}
  \label{fig:fi}
\end{figure}

As we can see, explicit interaction $g_{ex}(\cdot)$ mainly  affects the phase of features (\ie $\theta_j$ and $\theta_k$), which can be approximately  considered  as the rotations in the complex vector space, while implicit interaction $g_{im}(\cdot)$ performs a 
parallelogram-like transformation in the complex vector space, which mainly affects the modulus instead of the phase (due to the limits in first quadrant). 
By integrating both implicit and explicit feature interactions, our approach can model the effect in both \emph{phase} and \emph{modulus}, thus leading to an improved capacity due to mutual enhancement. 
Figure~\ref{fig:fi} presents a  geometric interpretation of explicit and implicit feature interactions.

To further understand the explicit interaction, we present an illustrative example with a simple interaction $\tilde{\bm e}_1^{0.33} \odot \tilde{\bm e}_2^{0.25} $ with two feature vectors: $\tilde{\bm e}_1 = [-8, 1]^\top$ and $\tilde{\bm e}_2 = [-16,4]^\top$. With some mathematical computations, we can get the following representation: $\tilde{\bm e}_1^{0.33} \odot \tilde{\bm e}_2^{0.25} = \hat{\bm r} + i \hat{\bm p} = [-0.99, 1.41]^\top + i [3.85, 0]^\top$.

\begin{table}[!tp]
  \captionsetup{font={small}}
  \small
  \caption{Comparison of different CTR methods. ``Unified'' denotes the integrated learning of  explicit and implicit interactions, ``UR'' means ``Unrestricted'' indicating the elements of feature embeddings could be arbitrary real value, and ``SV'' means ``Single-vector'' indicating that each feature is represented by only one embedding vector.}
  \label{tab:cmp}
  \resizebox{\columnwidth}{!}{
  \begin{tabular}{@{}lcccccc@{}}
  \toprule
  \multirow{2}{*}{Methods} & \multicolumn{3}{c}{Feature Interaction} & \multicolumn{2}{c}{Embedding} & \multirow{2}{*}{Complexity}
  \\ \cmidrule(l){2-4} \cmidrule(l){5-6}
            & High-order & Adaptive & Unified & UR & SV           \\ \midrule
  FFM~\cite{juan2016field}  & \textcolor{purple}{\XSolidBrush} & \textcolor{purple}{\XSolidBrush} & \textcolor{purple}{\XSolidBrush} & \textcolor{teal}{\CheckmarkBold} & \textcolor{purple}{\XSolidBrush} & $O(m^2d)$\\
  FmFM~\cite{sun2021fm2}    & \textcolor{purple}{\XSolidBrush} & \textcolor{purple}{\XSolidBrush} & \textcolor{purple}{\XSolidBrush} & \textcolor{teal}{\CheckmarkBold} & \textcolor{teal}{\CheckmarkBold} & $O(m^2d^2)$\\ 
  DCNV2~\cite{wang2021dcn}   & \textcolor{teal}{\CheckmarkBold} & \textcolor{purple}{\XSolidBrush} & \textcolor{purple}{\XSolidBrush} & \textcolor{teal}{\CheckmarkBold} & \textcolor{teal}{\CheckmarkBold} & $O(m^2d^2L + mHd + H^2T)$\\
  AFN+~\cite{cheng2020adaptive} & \textcolor{teal}{\CheckmarkBold} & \textcolor{teal}{\CheckmarkBold} & \textcolor{purple}{\XSolidBrush} & \textcolor{purple}{\XSolidBrush} & \textcolor{purple}{\XSolidBrush} & $O(mdK + KHd + mHd + H^2T)$\\ 
  EulerNet (ours) & \textcolor{teal}{\CheckmarkBold}  & \textcolor{teal}{\CheckmarkBold} & \textcolor{teal}{\CheckmarkBold} & \textcolor{teal}{\CheckmarkBold} & \textcolor{teal}{\CheckmarkBold} & $O(mnd^2 + n^2d^2L)$\\ \bottomrule
  \end{tabular}
  }
  \end{table}

\subsubsection{Novelty and differences.}
In Table~\ref{tab:cmp}, we compare our approach with existing feature interaction methods. 
To the best of our knowledge, it is the first attempt that adaptively captures arbitrary-order feature interactions in the complex vector space.
Although AFN+~\cite{cheng2020adaptive} leverages the LNN~\cite{hines1996logarithmic} to learn arbitrary-order feature interactions adaptively, it constrains the feature representations to positive real vectors. This approach not only degrades the model performance, but also requires additional feature embeddings for implicit interactions.
Furthermore, most studies~\cite{wang2021dcn, lian2018xdeepfm, guo2017deepfm,yu2020deep} model the explicit and implicit interactions in different architectures  and seldom integrate them in a joint approach. 
As a comparison, EulerNet is more \emph{general, unified} in integrating the modeling of implicit and explicit feature interactions, via the enhanced Euler interaction layer in Section~\ref{sec:implicit}.  In general, our approach provides a more capable solution to model complicated feature interactions. 



\subsubsection{Complexity Analysis}
We also compare the time complexities of different CTR methods in Table~\ref{tab:cmp}.
For ease of analysis, we assume that the hidden size of different components is set to  the same  number.
Specially, $m$ is the number of feature fields, $d$ is the embedding dimension, $L$ and $T$ are the layer depth of the explicit and implicit component respectively, $H$ is the hidden size of MLP, $K$ is the number of logarithmic neurons of AFN+~\cite{cheng2020adaptive}, and $n$ is the number of order vectors of EulerNet.
Note that $K$ is much larger than $m\cdot d$, leading to a very high complexity of AFN+~\cite{cheng2020adaptive}.
In contrast, $n$ is very small, which can be set to $m$ in practice.
The complexity of EulerNet for a training instance can be estimated as $O(m^2d^2L)$, which is comparable to mainstream efficient methods such as FmFM~\cite{sun2021fm2} and DCNV2~\cite{wang2021dcn} (See Table~\ref{tab:overall} for experimental analysis).
\section{EXPERIMENTS}
We conduct extensive experiments to show the effectiveness of EulerNet, and analyze the effect of each learning component in it.

\subsection{Experimental Settings}
We introduce the experimental settings, including the datasets,  baseline approaches, and the details of hyper-parameters.

\subsubsection{Datasets}
We utilize three real world datasets in our experiments: Criteo\footnote{http://labs.criteo.com/2014/02/kaggle-display-advertising-challenge-dataset}, Avazu\footnote{http://www.kaggle.com/c/avazu-ctr-prediction}, MovieLens-1M\footnote{https://grouplens.org/datasets/movielens}. 
Table~\ref{tab:datasets} summarizes the dataset statistics information.

$\bullet$ Criteo. The most popular CTR prediction benchmark dataset contains user logs over a period of 7 days. 

$\bullet$ Avazu. It contains user logs over a period of 7 days, which was used in the Avazu CTR prediction competition.

$\bullet$ MovieLens-1M. The most popular dataset for recommendation systems research.

  \begin{table}[!h]
    \centering
    \small
    \captionsetup{font={small}}
    \caption{The statistics of datasets.} 
    \label{tab:datasets}
    \begin{tabular}{c|ccc}
      \toprule
      \textbf{Dataset}& $\#$ Features & $\#$ Fields & $\#$ Instances \\
      \hline \hline
      Criteo & 1.3M & 39 & 45M \\
      Avazu & 1.5M & 23 & 40M \\
      MovieLens-1M & 13k & 7 & 740K\\
    \bottomrule
  \end{tabular}
  \end{table}





\subsubsection{Compared Models.}
We compare EulerNet with state-of-the-art methods in CTR prediction task, including:

$\bullet$ FwFM~\cite{pan2018field} improves FM by considering field information and uses field-specific weights to capture the field-wise relationship.

$\bullet$ FmFM~\cite{sun2021fm2} replaces the field scalar weight in FwFM with a kernel matrix, allowing for modeling more informative interactions.

$\bullet$ DeepFM~\cite{guo2017deepfm} uses FM to model the second-order interactions, and incorporates DNNs to model the high-order interactions.

$\bullet$ DeepIM~\cite{yu2020deep} utilizes Newton's identity to implement high-order FM, and incorporate implicit interactions via an MLP.

$\bullet$ xDeepFM~\cite{lian2018xdeepfm} encodes high-order interactions into multiple feature maps and combine an MLP to model implicit interactions.

$\bullet$ DCNV2~\cite{wang2021dcn} takes the kernel product of concatenated feature vectors to model high-order interactions and combine an MLP to model implicit interactions.

$\bullet$ FiBiNet~\cite{huang2019fibinet} uses the bilinear operation to  model pair-wise interactions and uses SENet~\cite{hu2018squeeze} to capture the 
 feature importance.

$\bullet$ AutoInt~\cite{song2019autoint} uses the self-attention mechanism to learn high-order interactions.
AutoInt+ improves it by combining an MLP.

$\bullet$ FiGNN~\cite{li2019fi} represents  the features into a full-connected graph, and uses gated GNNs to model the high-order feature interactions.


$\bullet$ AFN~\cite{cheng2020adaptive} encodes features into a logarithmic space to adaptively learn the arbitrary-order feature interactions.
AFN+ improves the base model by using an MLP to model implicit interactions.

The above models we compared in our experiments have covered different types of feature interaction methods. 
FwFM and FmFM are shallow models that only model the second-order explicit interactions.
DeepFM, DeepIM, xDeepFM and DCNV2 are ensemble methods that learn both the explicit interactions by an empirically designed component and implicit interactions by an MLP.
FiBiNet, AutoInt, and FiGNN have the ability to learn the importance of feature interactions.
AFN encodes features into a logarithmic space to adaptively learn the arbitrary-order feature interactions.
Different from them, our proposed EulerNet represents the features in a complex vector space, in which the exponential computation can be simplified as linear computation,
making it possible to adaptively learn the arbitrary-order feature interactions in an efficient way.

\subsubsection{Implementation Details} \label{sec: impd}
All methods are implemented in Pytorch \cite{paszke2019pytorch}.
The size of feature embedding is 16.
The learning rate is in \{1e-3, 1e-4, 1e-5\}. 
The $L_2$ penalty weight is in \{1e-3, 1e-5, 1e-7\}. 
The batch size is 1024.
The training optimizer is Adam~\cite{kingma2014adam}.
The hidden layer of MLP component is $400 \times 400 \times 400$ and the dropout rate is 0.1.
For DeepIM, the interaction order is in \{2, 3, 4\}.
For xDeepFM, the depth of CIN is in \{1, 2, 3, 4, 5\} and the hidden size is in \{100, 200, 400\}.
For DCNV2, the depth of CrossNet is in \{1, 2, 3, 4, 5\}.
For FiGNN, the graph interaction step is in \{1, 2, 3, 4, 5\}.
For AutoInt+, the depth, number of head and attention size is 2, 2, 40 respectively.
For AFN, the number of logarithmic neurons is in \{40, 400, 800, 1000\}. 
For EulerNet, the number of Euler interaction layer is in \{1, 2, 3, 4, 5\}, and the number of order vectors is set as \{7, 23, 39\} for MovieLens-1M, Avazu and Criteo datasets respectively.
Our implementation is also available at RecBole~\cite{zhao2021recbole, zhao2022recbole}.

\subsection{Overall Performance}

\begin{table*}[!t]
  \centering
  \small
  \captionsetup{font={small}}
  \caption{Performance comparisons. A higher AUC or lower Logloss at 0.001-level is regarded significant, as stated in previous studies~\cite{song2019autoint, cheng2016wide, guo2017deepfm,wang2017deep}.}\label{tab:overall}
  \begin{tabular}{c|ccrr|ccrr|ccrr}
    \toprule
    \multirow{2}{*}{\textbf{Model}}&
    \multicolumn{4}{c|}{\textbf{Criteo}}&\multicolumn{4}{c|}{\textbf{Avazu}}&\multicolumn{4}{c}{\textbf{MovieLens-1M}}\\
    &AUC&LogLoss& Params & Latency & AUC & LogLoss & Params & Latency & AUC & LogLoss & Params &  Latency \\
    \hline\hline
    FwFM & 0.8104 & 0.4414 & 0.74 K & 6.71 ms & 0.7741 & 0.3835 & 0.25 K & 3.96 ms & 0.8815 & 0.3351 & 0.02 K & 1.12 ms\\
    FmFM & 0.8112 & 0.4408 & 0.39 M & 8.06 ms & 0.7744 & 0.3831 & 0.13 M & 4.32 ms & 0.8864 & 0.3295 & 0.01 M & 1.26 ms\\
    DeepFM & 0.8121 & 0.4401 & 0.57 M & 10.31 ms & 0.7830 & 0.3790 & 0.47 M & 6.94 ms &  0.8935 & 0.3230 & 0.36 M & 3.78 ms \\
    DeepIM & 0.8124 & 0.4397 & 0.57 M & 10.86 ms & 0.7838 & \underline{0.3779} & 0.47 M & 7.19 ms & 0.8927 & 0.3230 & 0.36 M & 4.20 ms\\
    xDeepFM & 0.8122 & 0.4407 & 2.44 M & 188.68 ms & 0.7821 & 0.3799 & 2.29 M & 78.01 ms & 0.8944 & 0.3235 & 0.52 M & 44.68 ms\\
    DCNV2 & \underline{0.8127} & \underline{0.4394} & 1.74 M & 16.39 ms & 0.7838 & 0.3782 & 0.87 M & 11.63 ms & 0.8946 & 0.3229 & 0.39 M & 4.03 ms\\
    FiBiNet & 0.8126 & 0.4415 & 9.82 M & 136.79 ms & 0.7837 & 0.3783 & 3.57 M & 32.08 ms & 0.8860 & 0.3291 & 0.87 M & 8.13 ms\\
    FiGNN & 0.8109 & 0.4412 & 0.08 M & 121.51 ms & 0.7830 & 0.3799 & 0.05 M & 46.31 ms &  0.8939 & 0.3232 & 0.01 M & 10.75 ms\\
    AutoInt+ & 0.8126 & 0.4396 & 3.80 M & 41.67 ms & 0.7838 & 0.3785 & 1.43 M & 18.52 ms & 0.8937 & 0.3288 & 1.17 M & 9.61 ms\\
    AFN+ & 0.8123 & 0.4396 & 19.46 M & 132.45 ms & \underline{0.7843} & 0.3785 & 14.19 M & 93.72 ms & \underline{0.8950} & \underline{0.3212} & 5.65 M & 84.18 ms\\
    \hline \hline
    EulerNet & \textbf{0.8137} & \textbf{0.4389} & 0.79 M & 13.51 ms & \textbf{0.7863} & \textbf{0.3769} & 0.27 M & 9.09 ms & \textbf{0.9008} & \textbf{0.3114} & 0.02 M & 2.69 ms\\
  \bottomrule
\end{tabular}
\label{tab:overall}
\end{table*}

We present the experimental results of different methods for CTR prediction in Table~\ref{tab:overall}, and have the following observations:

(1) Compared to the DNN-based methods, FwFM and FmFM perform worst due to the limited ability to only capture the second-order explicit feature interactions.

(2) Ensemble methods (\ie DeepFM~\cite{guo2017deepfm}, DeepIM~\cite{zhu2020ensembled}, DCNV2~\cite{wang2021dcn} and xDeepFM~\cite{lian2018xdeepfm}) achieve competitive performance across on all three datasets, which shows the effectiveness of integrating implicit feature interactions.

(3) For the feature importance learning methods (\ie FiBiNet~\cite{huang2019fibinet}, FiGNN~\cite{li2019fi} and AutoInt+~\cite{song2019autoint}), their performance largely varies across different datasets.
AutoInt+ performs very well on all three datasets, while FiGNN and FiBiNet lose the advantage on the Criteo and MovieLens-1M datasets respectively.
This indicates that the self-attention mechanism is more capable in modeling high-order feature interactions.

(4) AFN+  outperforms all the other baseline methods on the Avazu and MovieLens-1M datasets, demonstrating the effectiveness of adaptively  learning the arbitrary-order feature interactions in the CTR prediction task.
However, its advantage on the Criteo dataset is small. This may be caused by the restriction of  positive values assigned in the feature embeddings, which hinders the representation capability of the model.

(5) Our proposed EulerNet consistently performs better than all of the compared methods. It shows the effectiveness of encoding the features into the complex vectors via Euler’s formula and conducting transformations in the polar space.

As for the model efficiency, we can see that the latency of FwFM, FmFM, DeepFM, DeepIM and DCNV2 are relatively small. 
They are more efficient due to the simple architecture and fewer parameters learned in the model.
For AutoInt, FiGNN, FiBiNet and xDeepFM, the latency of them is much larger due to the complex model architecture or the complicated training strategy.
For AFN+, due to the limited feature representation space, \ie  only positive feature embeddings can be learned in logarithmic space transformation, it requires a large amount of parameters for retaining the performance. 
This makes it impractical in the industrial scenarios.
In contrast, the latency of EulerNet is much less than AFN+ (\ie under $10.2\%$) and it is comparable to many efficient methods such as DeepFM and DeepIM. 
With the highest accuracy and lower complexity, EulerNet has a great potential to be applied into large-scale industrial recommender systems.

\subsection{Experimental Analysis}
We conduct experiments to investigate the  interaction orders learned in EulerNet, and then visualize the learned feature representations to show their correlation with the feature importance.

\subsubsection{Arbitrary-Order Learning Analysis} 
\begin{figure}[!h]
  \centering
  \captionsetup{font={small}}
  \subcaptionbox{MovieLens-1M}{
    \includegraphics[width=0.467\linewidth]{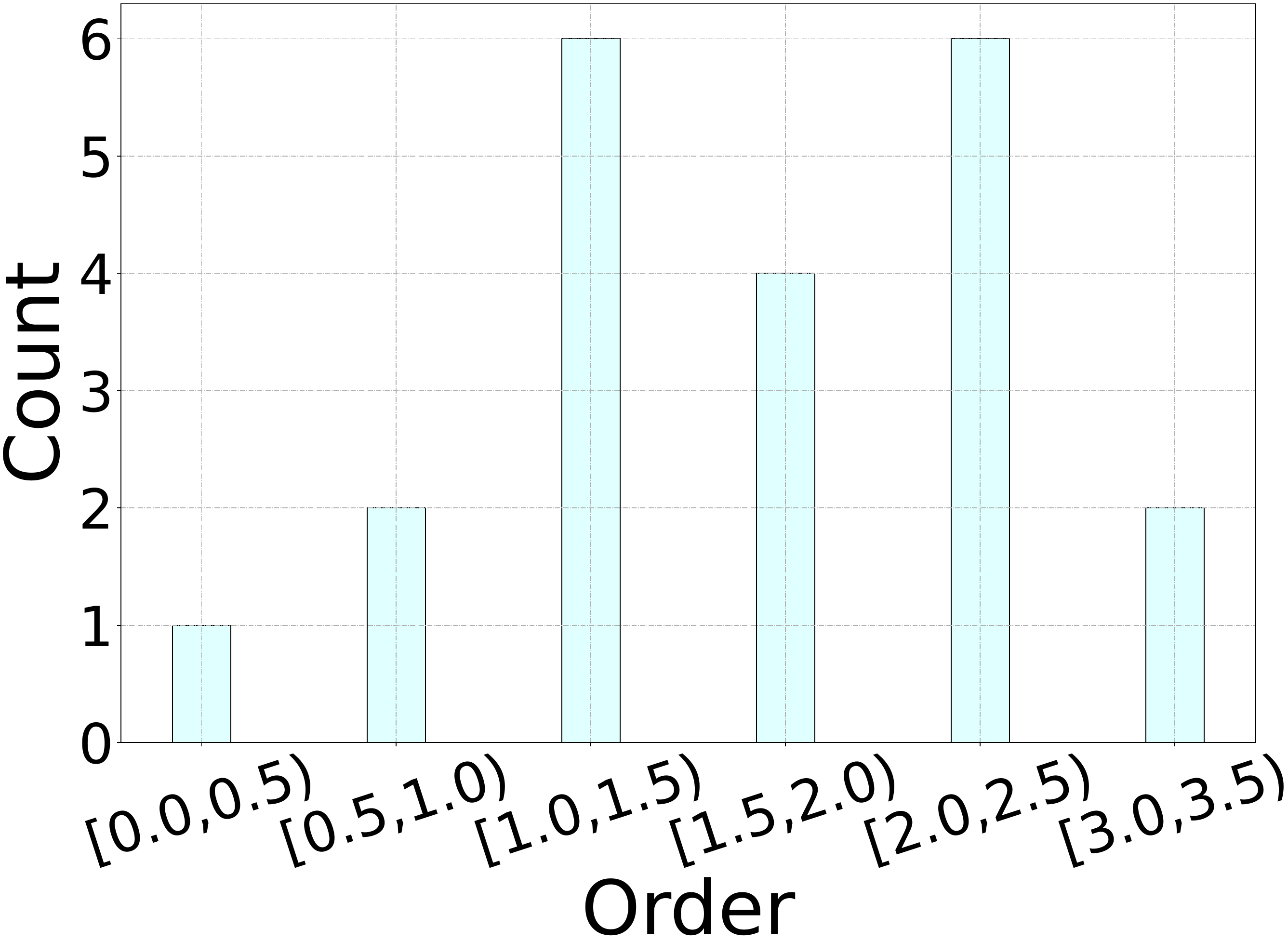}
  }
  \subcaptionbox{Avazu}{
    \includegraphics[width=0.477\linewidth]{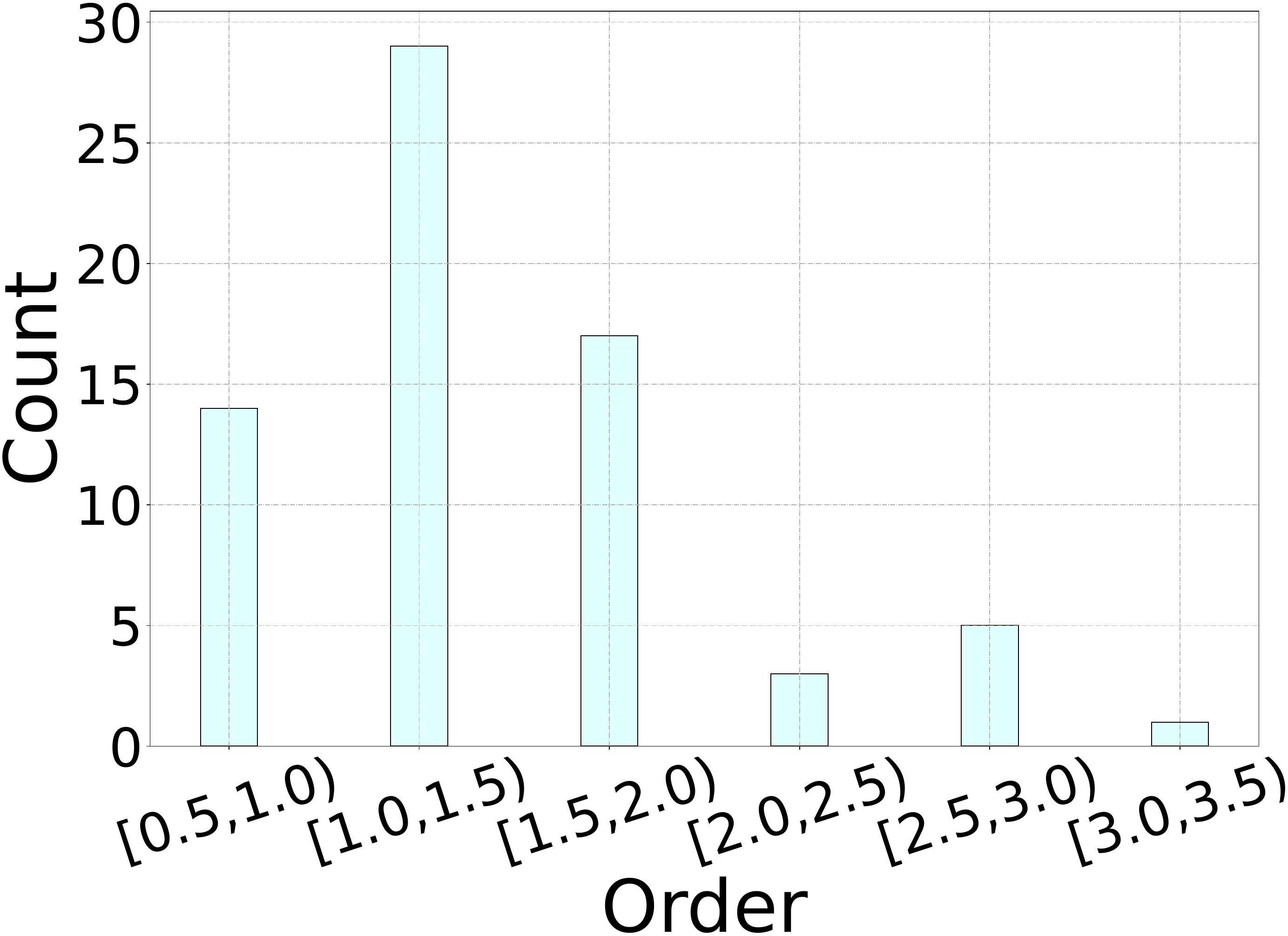}
  }
  \captionsetup{font={small}}
  \caption{The statistics of the interaction orders learned in EulerNet.}
  \label{fig:quant}
\end{figure}

Learning effective high-order feature interactions is very crucial in the CTR prediction task. 
To verify the orders learned in EulerNet, we visualize the learned feature interaction orders (\ie the total order of each learnable coefficient vector $\bm \alpha_k$ in Eq.~\eqref{eq:psi}) in the explicit feature interaction component. 
Note that the orders adaptively learned in our model can be arbitrary values in $[-\infty, +\infty]$, we cluster them  by setting the interval to 0.5 for better presentation.
As shown in Figure~\ref{fig:quant}, we can see that our model not only learns the integer-order feature interactions, but also can adaptively learn the fractional-order feature interactions in a fine-grained way.
Specially, the feature interaction orders learned in our model varies from $[0, 3.5]$ on MovieLens-1M dataset and $[0.5, 3.5]$ on Avazu dataset. 
Fine-grained feature interaction learning can improve the capability of our model and enable it to capture more effective information for CTR prediction.

\subsubsection{Verification on Synthetic Dataset}
Since it can not identify the ground-truth of meaningful feature interactions in real-world public datasets, we further conduct an experiment using synthetic data to verify the degree of coincidence with the learned orders in EulerNet. 
The synthetic dataset consists of 1 million synthesized click-through records with 7 fields ($F = [f_1, f_2, ..., f_7]$) that simulate real click-through records.
Each field is independently created and contains one thousand features, and each feature is assigned a probability that affects the likelihood of a click-through event occurring.
For a given record $x_i = [p_1, p_2, ..., p_7]$, its label $y_i$ is generated by sampling from a probability distribution that is pre-defined by one of the patterns, \ie $R\in [R_1, R_2,R_3]$ (See Table~\ref{tab:pat}).
We compare the interaction orders learned in the explicit feature interaction component of AFN+ and EulerNet, and utilize fitting deviation to evaluate the difference between the orders in different learning algorithms and the ground-truth pattern $R$. 
From Table~\ref{tab:pat}, we can see that the deviation in EulerNet is much smaller than AFN+, demonstrating that EulerNet has the ability to adaptively learn more meaningful feature interactions. 

\begin{figure}[!h]
  \centering
  \includegraphics[width=.8\linewidth]{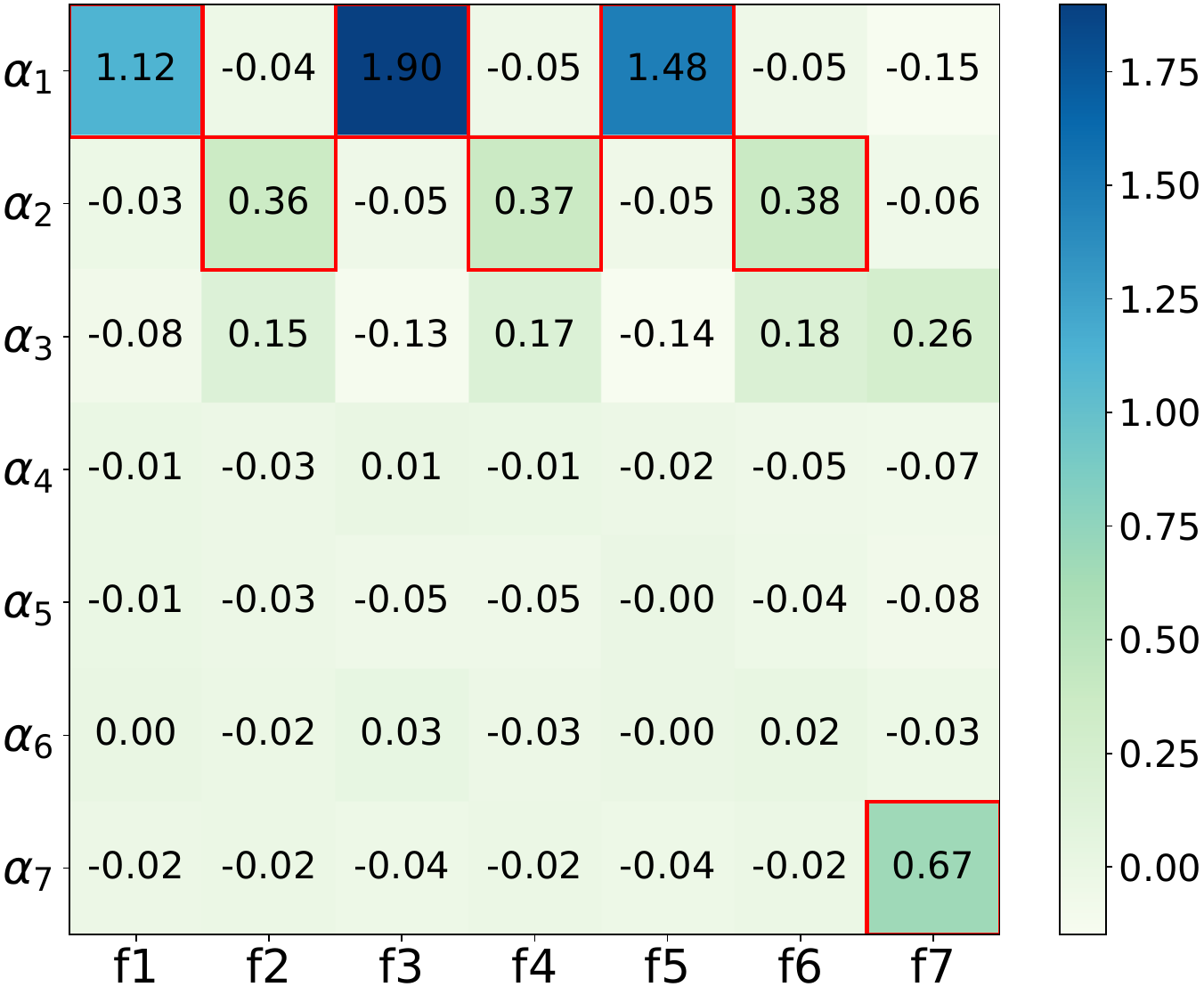}
  \captionsetup{font={small}}
  \caption{Visualization of the order vectors in EulerNet under the distribution pattern $R_3$, which is defined in Table~\ref{tab:pat}.}
  \label{fig:vis}
\end{figure}

Specifically, we present the order vectors of EulerNet after training on the synthetic dataset defined by the pattern $R_3$ in Figure~\ref{fig:vis}.
Different rows represent the explicit feature interactions learned by different order vectors (See Eq.~\eqref{eq:psi}). 
For example, the most important feature interactions learned by the order vector $\bm \alpha_1$ is $p_1^{1.12}p_3^{1.90}p_5^{1.48}$. 
We can see that the combinational feature interactions learned by a group of multi-order vectors $\{\bm \alpha_1, \bm \alpha_2, \bm \alpha_7\}$ (\ie $p_1^{1.12}p_3^{1.90}p_5^{1.48} + p_2^{0.36}p_4^{0.37}p_6^{0.38} + p_7^{0.67}$) are quite similar (average order deviation is 0.47) to the ground truth interaction pattern (\ie $R_3=\frac{1}{3}(p_1^{1.3}p_3^{2.2}p_5^{1.7} + p_2^{0.5}p_4^{0.5}p_6^{0.5} + p_7)$) in the data, showing the ability of EulerNet to learn the effective feature interactions.

\begin{table}[!h]
  \centering
  \small
  \captionsetup{font={small}}
  \caption{The pattern for creating the synthetic dataset and the deviation comparisons between different models.} 
  \label{tab:pat}
  \begin{tabular}{c|c|cc}
    \toprule
    \multirow{2}{*}{\textbf{Pattern}}& \multirow{2}{*}{\textbf{Formula}} & \multicolumn{2}{c}{\textbf{Deviation}} \\
    &&AFN+&EulerNet\\
    \hline \hline
    $R_1$ & $p_1^{0.3}p_3^{1.7}p_4^{2.5}$ & 0.6296 & 0.1141\\
    $R_2$ & $\frac{1}{2}(p_1^{1.7}p_2^{1.7} + p_3^{0.5}p_4^{2.5})$ & 2.4021 & 0.7481\\
    $R_3$ & $\frac{1}{3}(p_1^{1.3}p_3^{2.2}p_5^{1.7} + p_2^{0.5}p_4^{0.5}p_6^{0.5} + p_7)$ & 1.4732 & 0.4779\\
  \bottomrule
\end{tabular}
\end{table}

\subsubsection{Visualization of Feature Embeddings}
\begin{figure}[!h]
    \captionsetup{font={small}}
    {
    \begin{minipage}[t]{1.\linewidth}
        \centering
        \includegraphics[width=.93\textwidth]{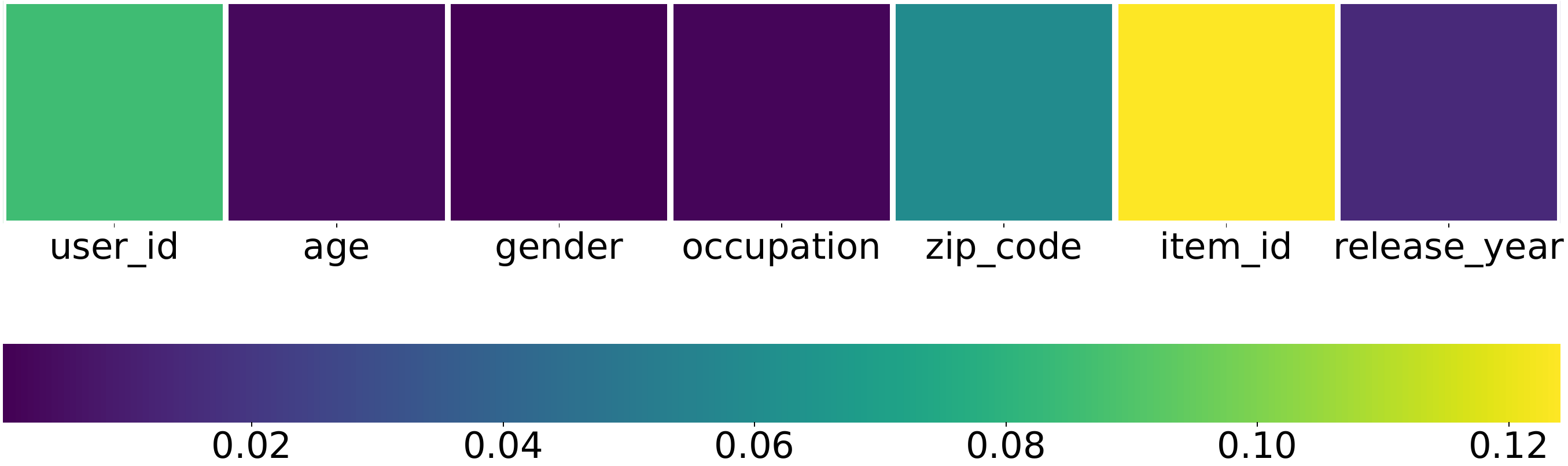}
    \end{minipage}
    \subcaption{Mutual information score between fields and label in MovieLens-1M.}
    }
    \vspace{0.5 em}
    {
    \begin{minipage}[t]{0.32\linewidth}
        \centering
        \includegraphics[width=1\textwidth]{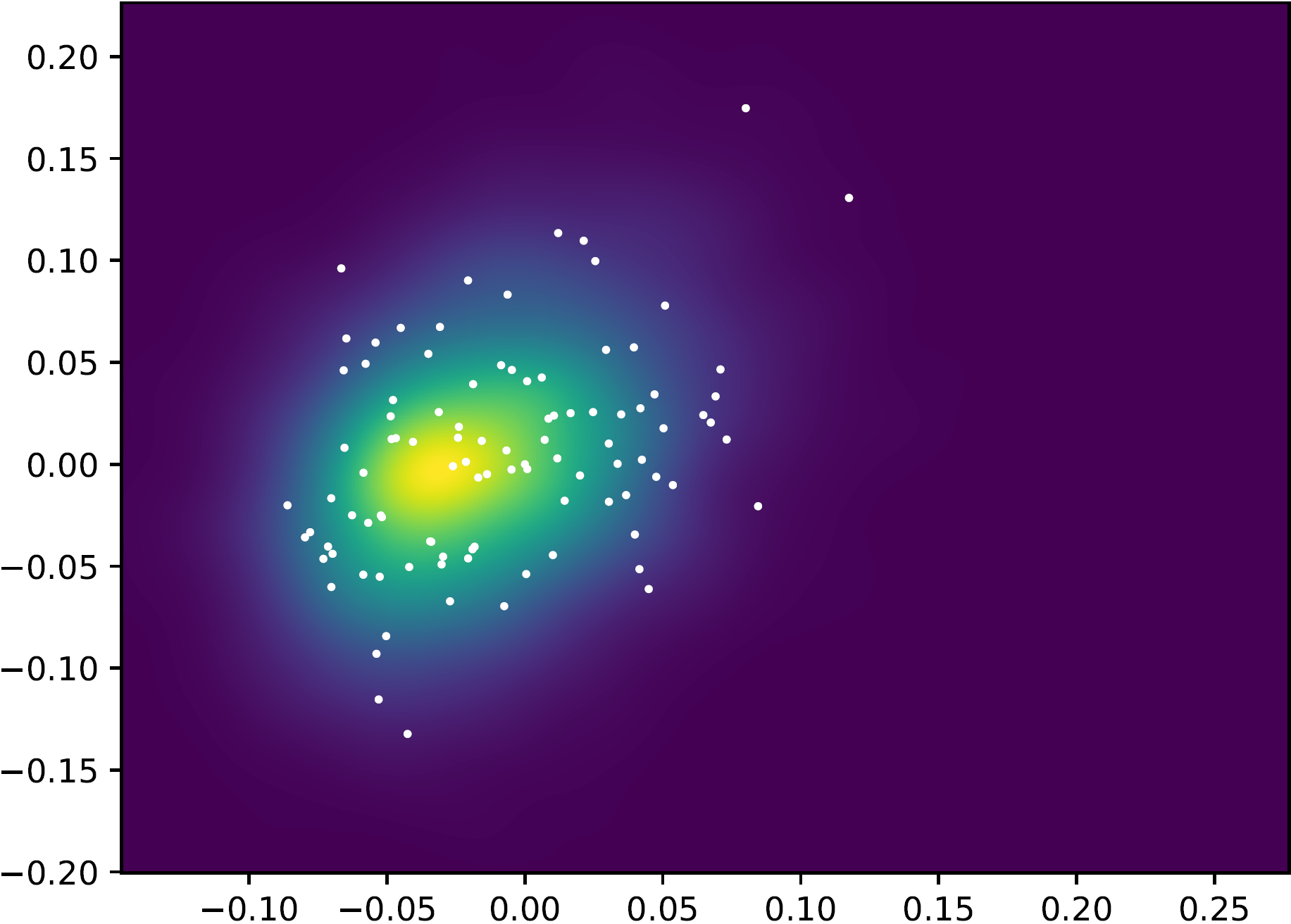}
        \subcaption*{user\_id} \label{fig:kde_ml1m_ncl}
    \end{minipage}
    \begin{minipage}[t]{0.32\linewidth}
		\centering
		\includegraphics[width=1\textwidth]{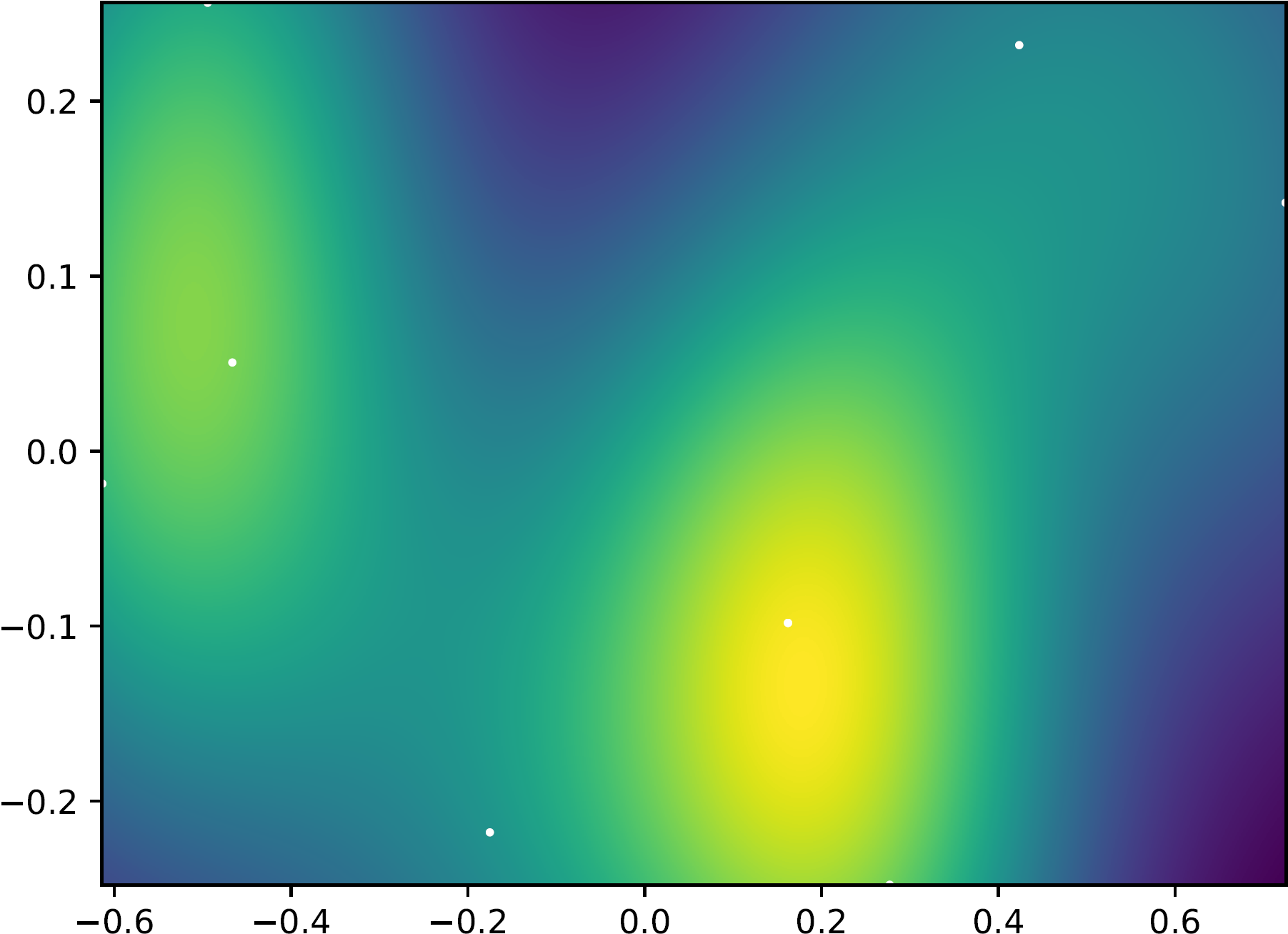}
		\subcaption*{age} \label{fig:kde_ml1m_lightgcn}
    \end{minipage}
    \begin{minipage}[t]{0.32\linewidth}
        \centering
        \includegraphics[width=1\textwidth]{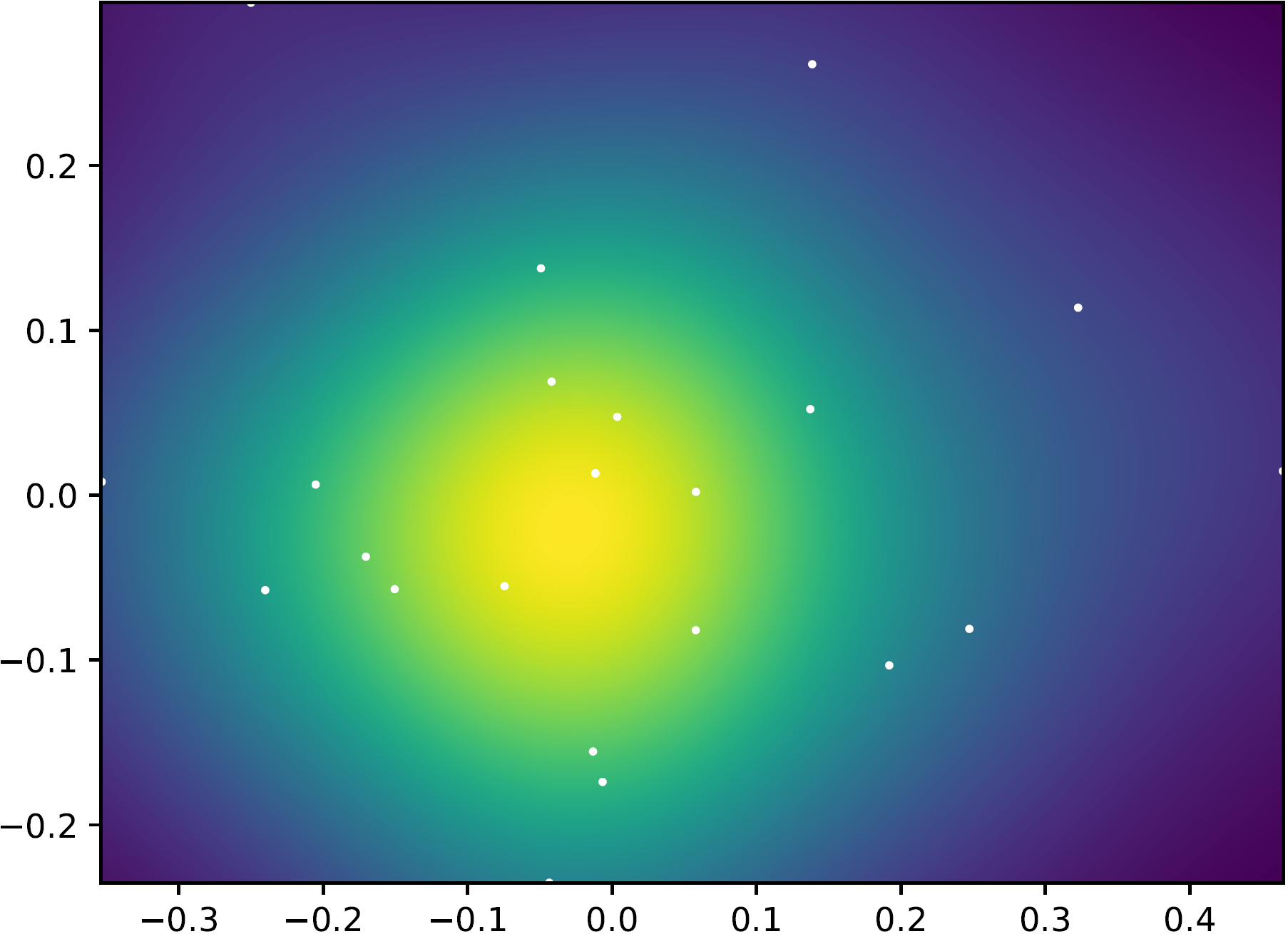}
        \subcaption*{occupation} \label{fig:opp}
    \end{minipage}
    
    \begin{minipage}[t]{0.32\linewidth}
        \centering
        \includegraphics[width=1\textwidth]{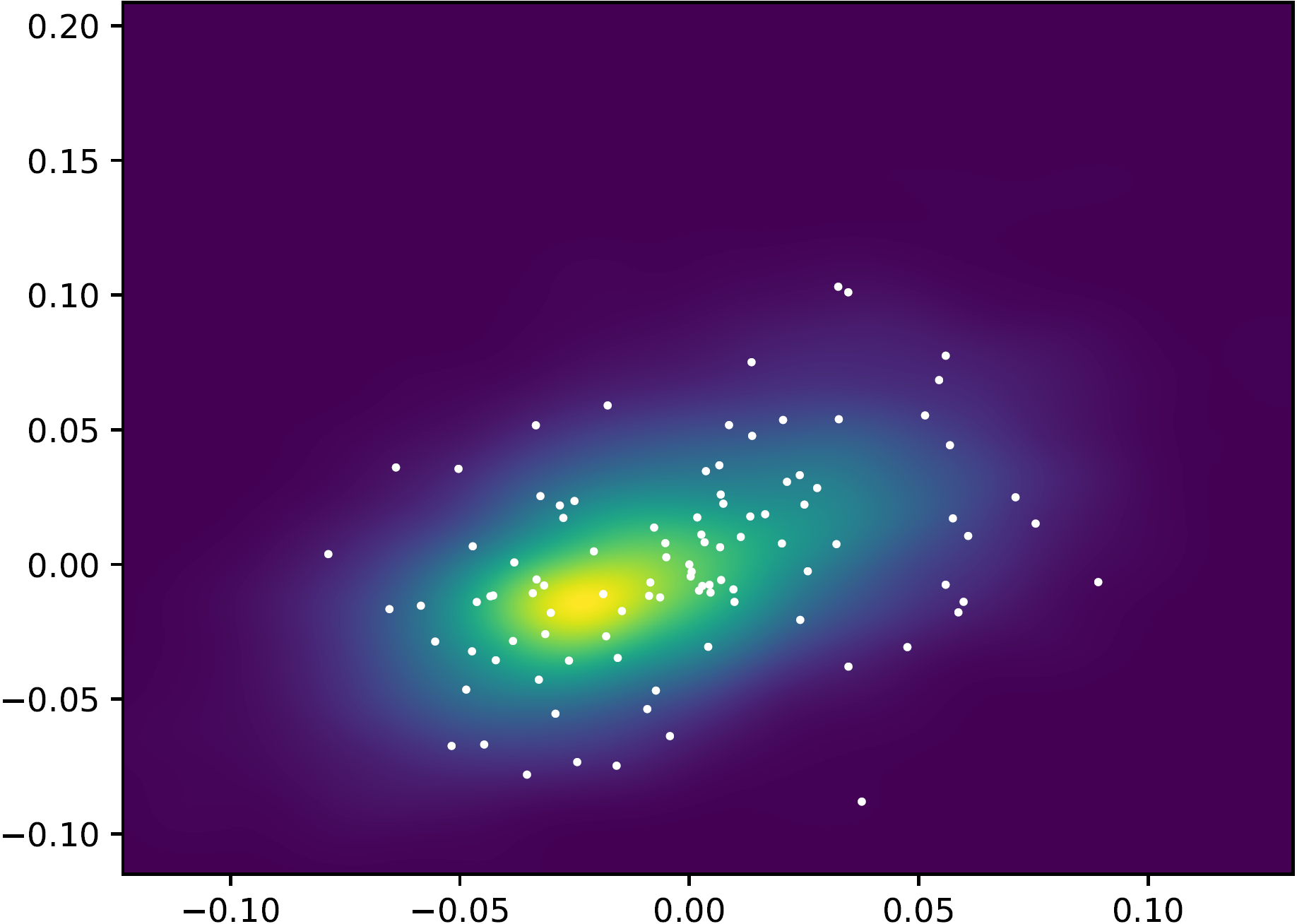}
        \subcaption*{zip\_code} \label{fig:kde_yelp_ncl}
    \end{minipage}
    \begin{minipage}[t]{0.32\linewidth}
        \centering
        \includegraphics[width=1\textwidth]{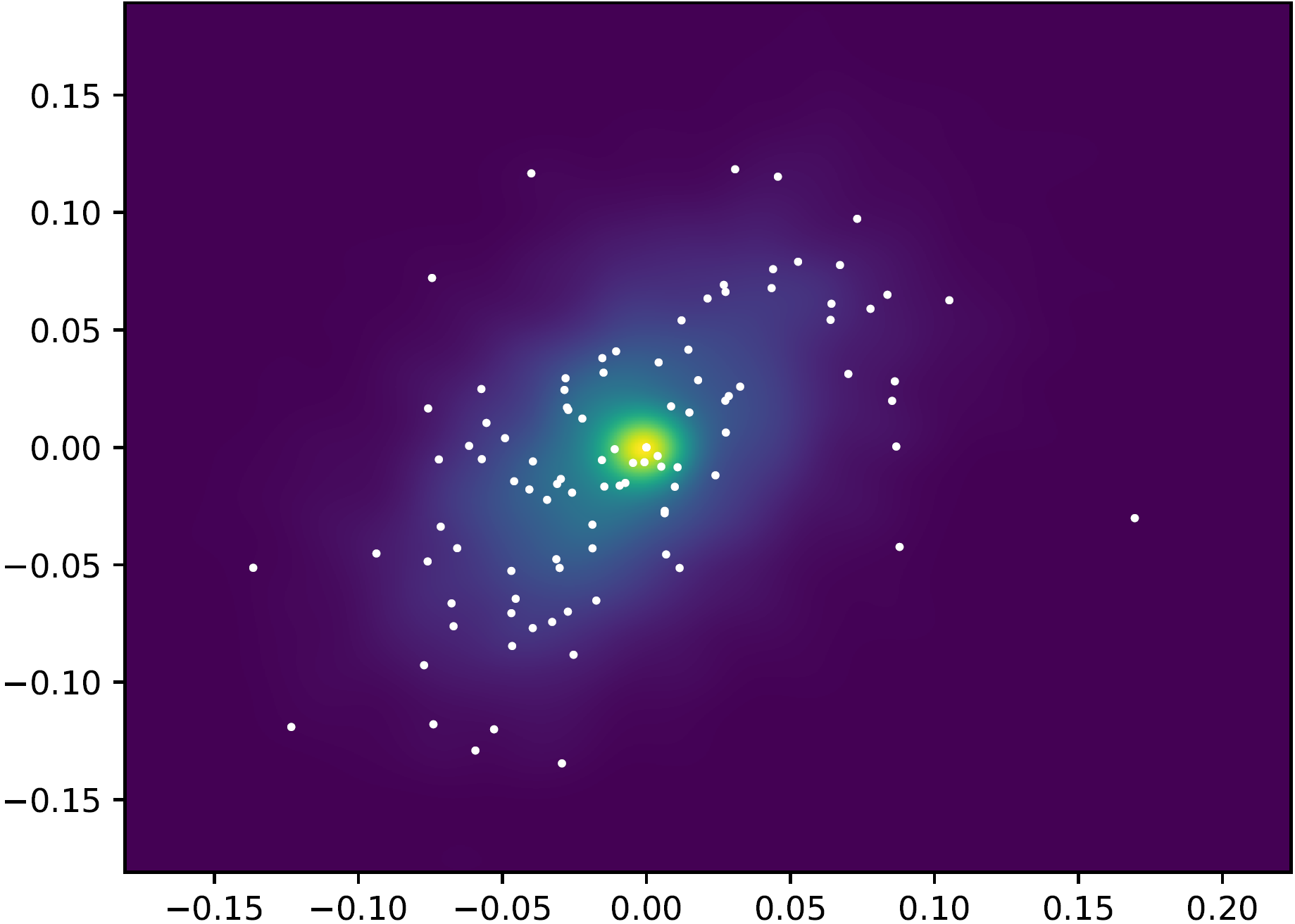}
        \subcaption*{item\_id} \label{fig:kde_yelp_ncl}
    \end{minipage}
    \begin{minipage}[t]{0.327\linewidth}
        \centering
        \includegraphics[width=1\textwidth]{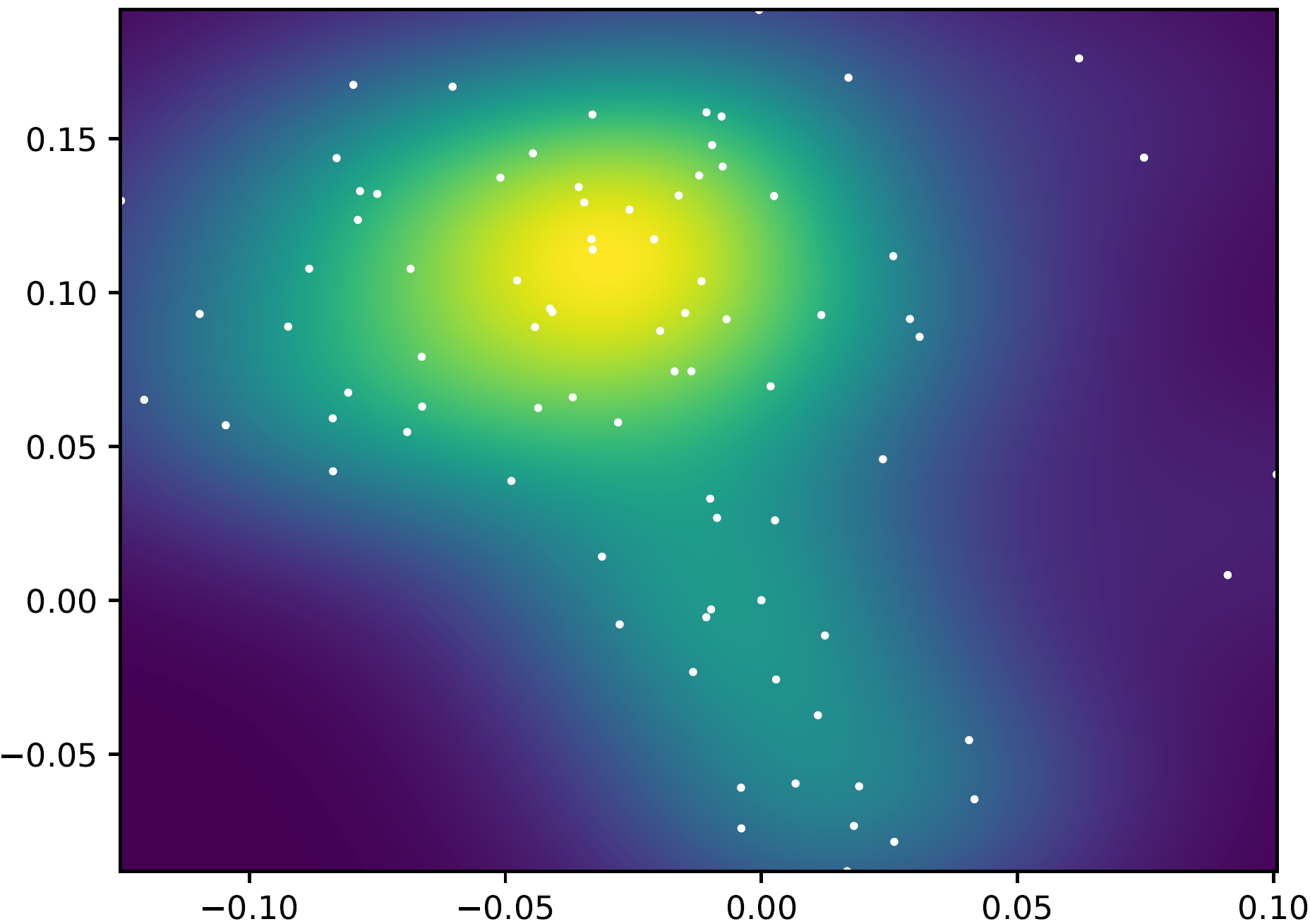}
        \subcaption*{release\_year} \label{fig:kde_yelp_ncl}
    \end{minipage}
    \subcaption{The distribution of feature embeddings in different feature fields.}
    }
    \caption{Visualization of the relationships between feature importance and the distribution of feature embeddings in EulerNet.}
    \label{fig:kde}
\end{figure}

EulerNet not only can adaptively learn the arbitrary-order feature interactions, but also have the ability to capture the importance of features. 
We visualize the learned feature embeddings in EulerNet and show its ability in learning the importance of features.
The heat map in Figure~\ref{fig:kde}(a) illustrates the mutual information scores between feature fields and labels on the MovieLens-1M dataset, which represents the strength of each field on the prediction results. 
We can observe that the fields \emph{item\_id}, \emph{user\_id} and \emph{zip\_code} have the strongest effect on the click results.
The distributions of feature embeddings are plotted with Gaussian kernel density estimation (KDE) in two-dimensional space in Figure~\ref{fig:kde}(b). 
The more dispersive the distribution of feature embeddings, the less influence of it has on the prediction results due to the low information quantity in the random varies.
It can be seen that for the important fields (\ie \emph{item\_id}, \emph{user\_id} and \emph{zip\_code}) in Figure~\ref{fig:kde}(a), the distribution of feature embeddings in Figure~\ref{fig:kde}(b) is more concentrated and has a smaller variance.
While for the fields with less importance (\ie \emph{age}, \emph{occupation} and \emph{release\_year}), they are chaotically distributed, and their variance is relatively large.
This indicates that the feature embeddings, which also represent the phase of the complex features as defined in Eq.~\eqref{eq-transform-1}, can reflect the feature importance to a certain extent.
In EulerNet, the phase of the complex features is effectively controlled by explicit feature interactions (See Section~\ref{sec:enhanceana}), which enables it to capture the meaningful feature relationship and improves the model capabilities.


\subsection{Ablation Study}\label{exp: alb}
We conduct ablation studies to explore the impact of each component or hyper-parameter on the model performance.

\subsubsection{Effect of Implicit and Explicit Feature Interactions}
EulerNet contains both the explicit and implicit interaction learning components. 
In order to investigate the impact of each interaction type, we conduct experiments on the two variants of EulerNet, termed as EulerNet$_{E}$ and EulerNet$_{I}$, in which the implicit and explicit learning parts are removed respectively.
As shown in Table~\ref{tab:fis}, we can see that the model performance has a decrease for both EulerNet$_{I}$ and EulerNet$_{E}$, showing the mutual complementary effects of them, which is consistent with the observations in Section~\ref{sec:enhanceana}. 
Besides, EulerNet$_{I}$ shows a larger decrease in performance than EulerNet$_{E}$ on the Avazu and MovieLens-1M datasets, but the decrease is smaller on the Criteo dataset,
showing both the implicit and explicit interactions are important for CTR prediction.

\begin{table}[!t]
  \centering
  \small
  \captionsetup{font={small}}
  \caption{Performance comparison between different interactions.} 
  \label{tab:fis}
  \begin{tabular}{c|c|ccc}
    \toprule
    \textbf{Dataset} & \textbf{Metric} & EulerNet & EulerNet$_E$ & EulerNet$_I$ \\
    \hline \hline
    \multirow{2}{*}{{Criteo}} & AUC & 0.8137 & 0.8117 & 0.8124\\
    & Decrease  & - & $-0.25\%$ & $-0.16\%$\\
    \hline
    \multirow{2}{*}{{Avazu}} & AUC & 0.7863 & 0.7847 & 0.7840\\
    & Decrease  & - & $-0.20\%$ & $-0.29\%$\\
    \hline
    \multirow{2}{*}{{MovieLens-1M}} & AUC & 0.9008 & 0.8988 & 0.8966\\
    & Decrease  & - & $-0.22\%$ & $-0.47\%$\\
  \bottomrule
\end{tabular}
\end{table}

\subsubsection{Impact of the Interaction Layer Number}
\begin{figure}[!t]
  \centering
  \captionsetup{font={small}}
  \subcaptionbox{AUC}{
    \includegraphics[width=0.477\linewidth]{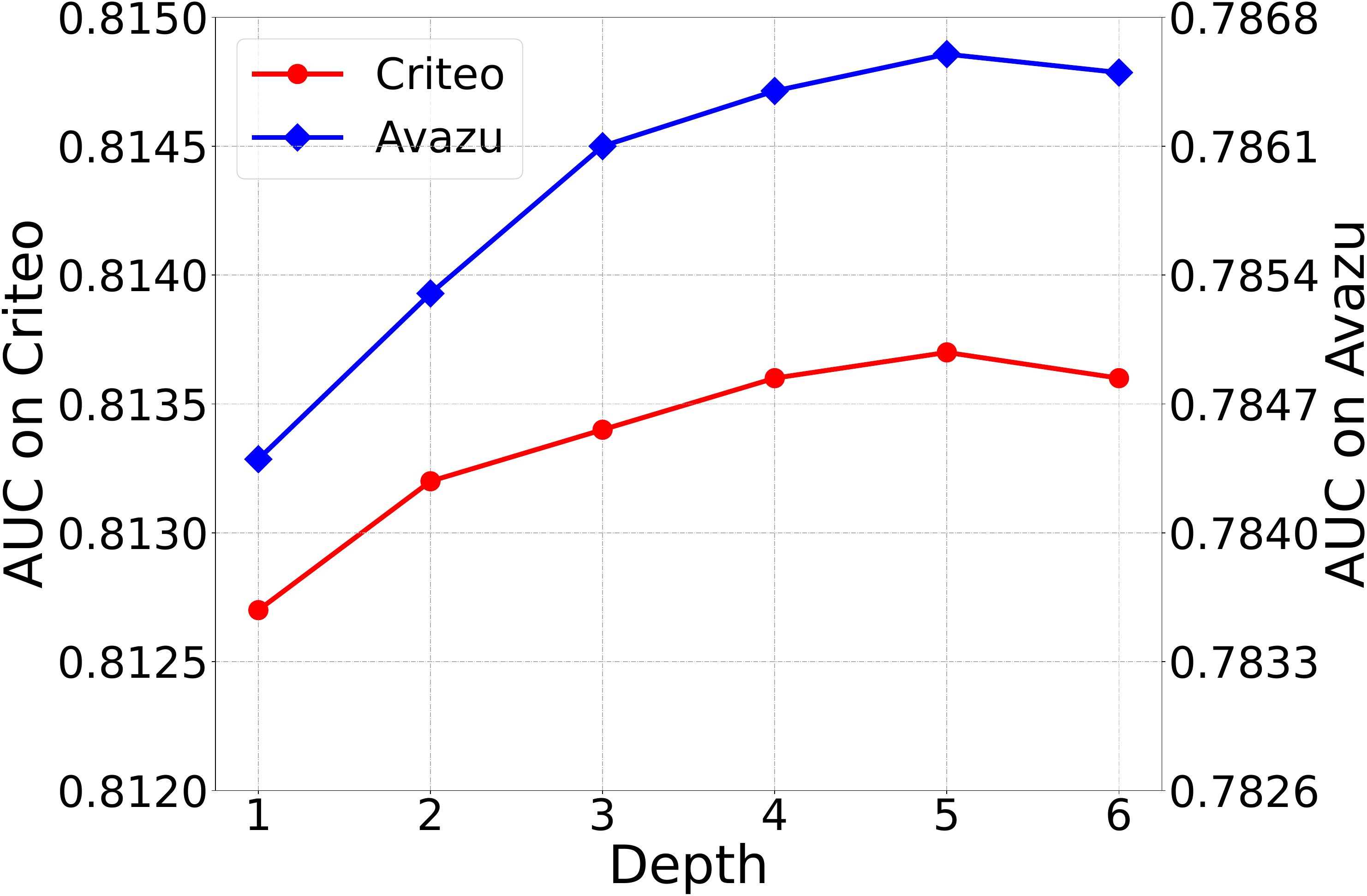}
  }
  \subcaptionbox{LogLoss}{
    \includegraphics[width=0.477\linewidth]{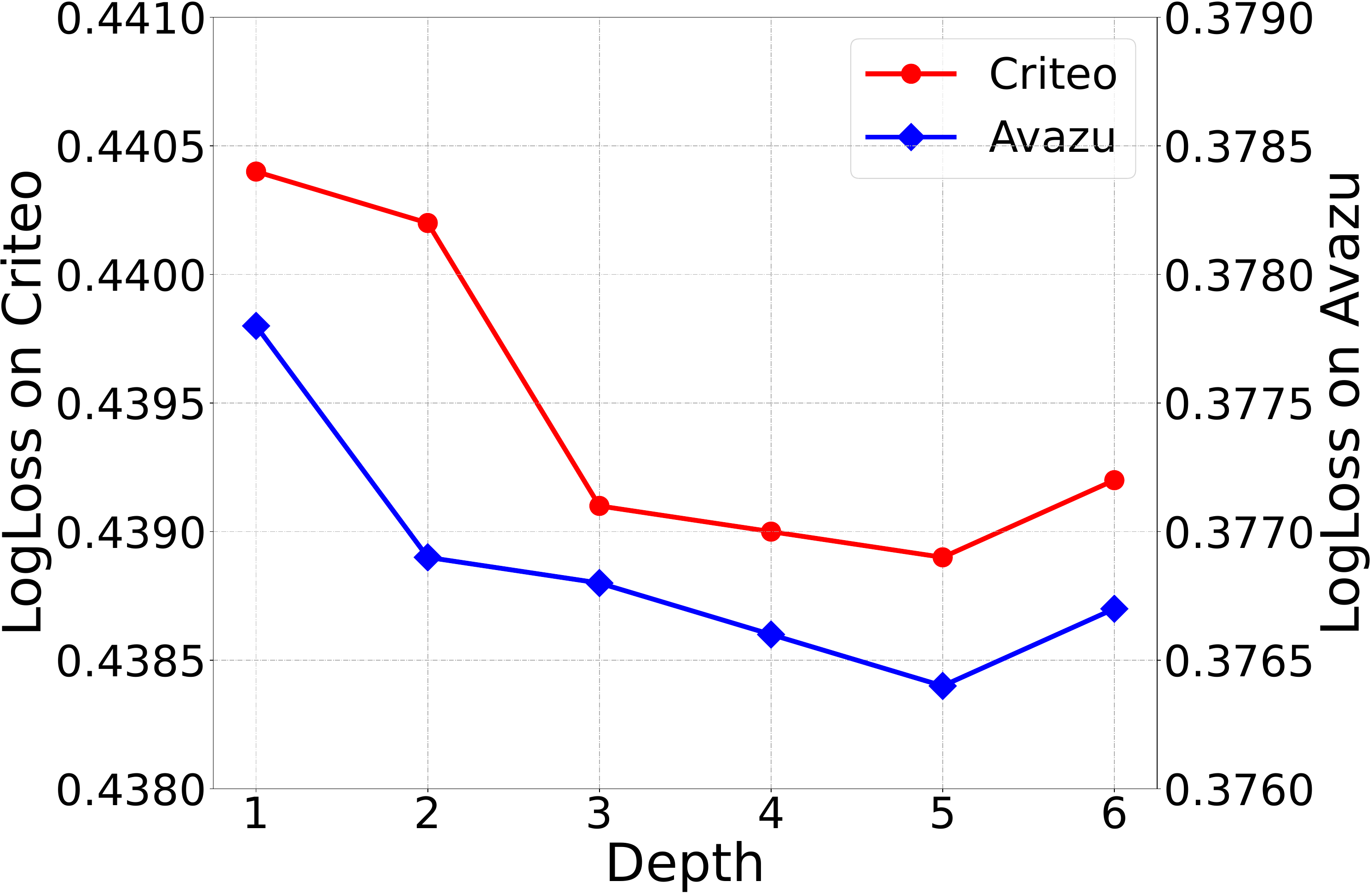}
  }
  \captionsetup{font={small}}
  \caption{Impact of the interaction layer number.}
  \label{fig:layer}
\end{figure}

EulerNet is designed by stacking the key structure of the Euler interaction layer.
We study the impact of the Euler interaction layer number, which reflects the intricacy of feature interactions, on the model performance.
As shown in Figure~\ref{fig:layer}, we can observe that the performance of EulerNet increases as the number of layers increases. 
EulerNet achieves the best model performance with 5 interaction layers. 
When the number of layers exceeds 5, the model performance decreases due to the overfitting issue caused by incorporating more parameters.

\subsubsection{Impact of the Number of Order Vectors.}
\begin{figure}[!t]
  \centering
  \captionsetup{font={small}}
  \subcaptionbox{AUC}{
    \includegraphics[width=0.477\linewidth]{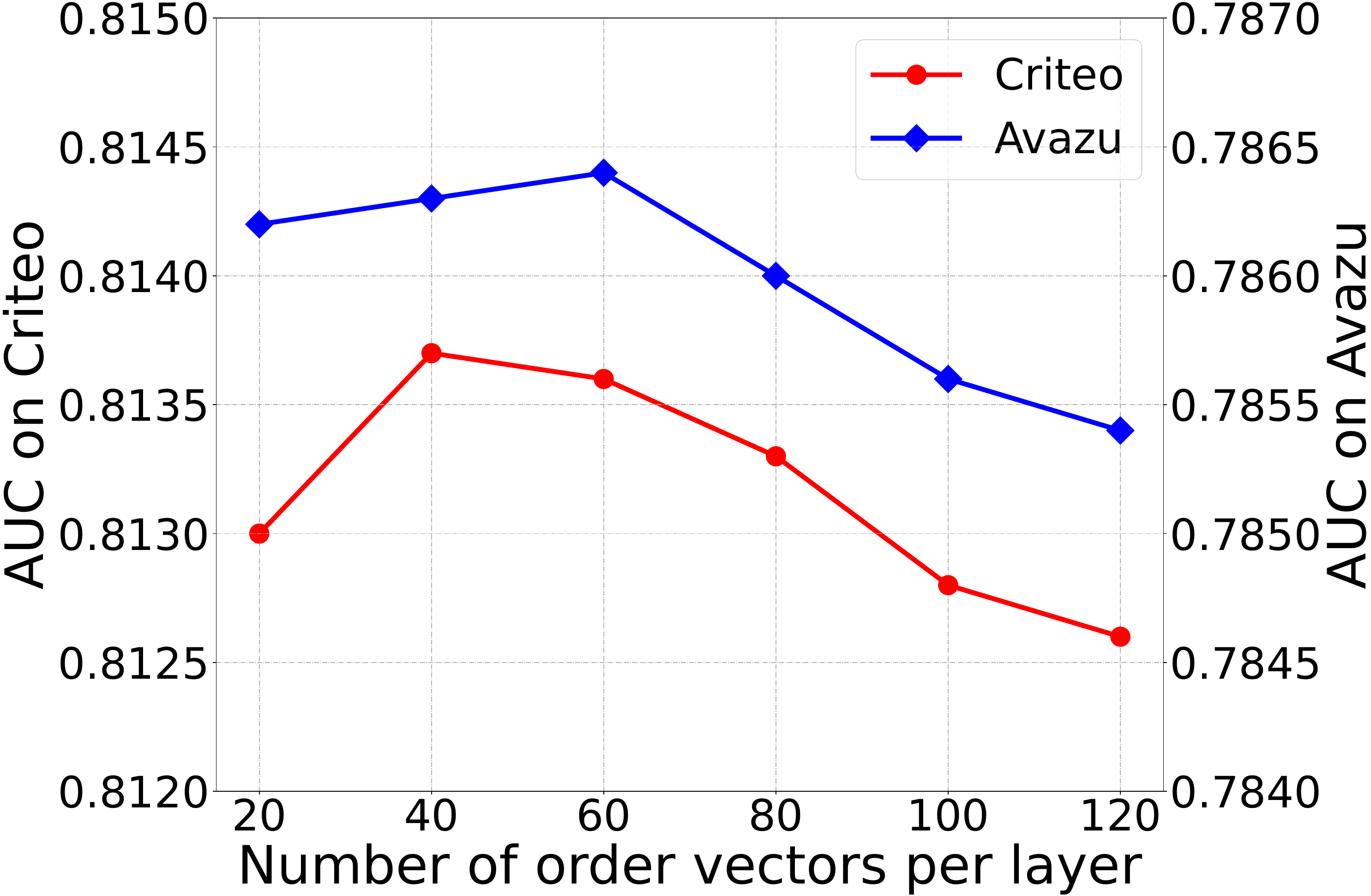}
  }
  \subcaptionbox{LogLoss}{
    \includegraphics[width=0.477\linewidth]{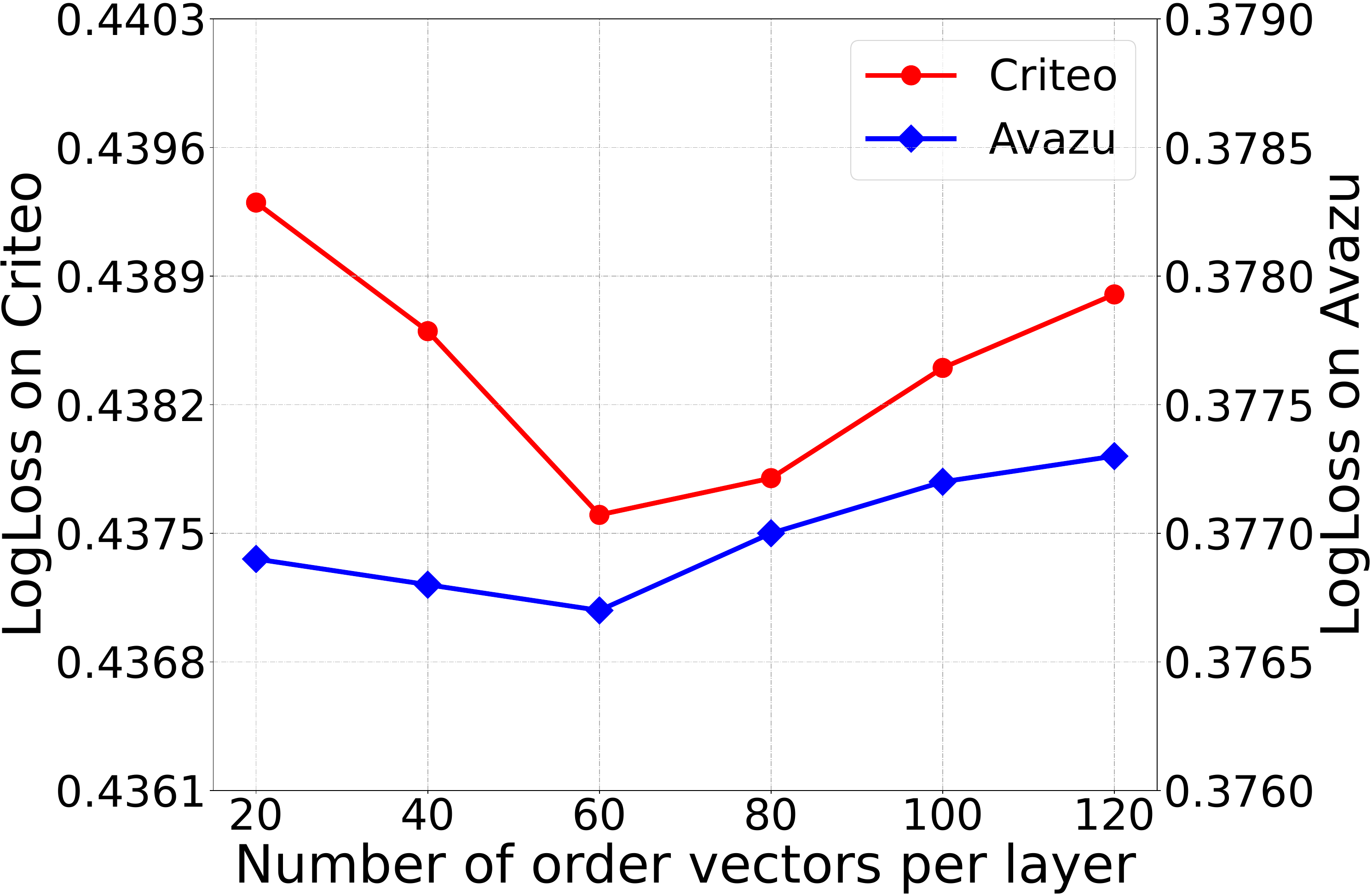}
  }
  \captionsetup{font={small}}
  \caption{Impact of the number of order vectors.}
  \label{fig:neurons}
\end{figure}

As introduced in Section~\ref{sec:eulerint}, we use multiple order vectors to adaptively learn the arbitrary-order feature interactions.
The number of order vectors is denoted as $n$ (See Eq.~\eqref{eq:group}), which controls the number of explicit feature interactions in each Euler interaction layer.
As illustrated in Figure~\ref{fig:neurons}, the performance of EulerNet on the Avazu dataset increases as the number of order vectors increases from 20 to 60.
Whereas on the Criteo dataset, EulerNet achieves the best performance as the number of order vectors increases to 40.
However, the model performance decreases when adding more order vectors.
This indicates that including too many feature combinations in the multi-order transformation may incorporate the useless feature interactions that hurt the model performance.

\section{Related work}

\paratitle{Explicit Feature Interaction Learning.}
This line of research explicitly enumerates feature combinations and uses vector operations such as inner product to capture their relationships. 
Early CTR models~\cite{richardson2007predicting, chang2010training, he2014practical,chen2016xgboost} mainly relied on manually designing feature combinations with simple architectures.
For example, FM~\cite{rendle2010factorization} assigns an embedding vector to each feature that mainly captures second-order interactions.
Inspired by FM, many variants of factorization machines have been proposed~\cite{cheng2014gradient, juan2016field, hong2019interaction, xu2020learning, lu2021dual}.
Among them, FFM~\cite{juan2016field} assigns multiple embeddings to explicitly model field-wise feature interactions.
Besides, FwFM \cite{pan2018field} and FmFM \cite{sun2021fm2} are proposed to model the field information to improve FM in a parameter-efficient way.
These factorization based methods  mainly model second-order interactions, which severely limits their performance.
To capture more effective feature interactions, 
xDeepFM~\cite{lian2018xdeepfm} proposes the CIN to model the high-order feature interactions by incorporating lots of learnable parameters. Besides, DCNV2~\cite{wang2021dcn} proposes the CrossNet to capture the high-order feature interactions in an efficient way.
Although these methods leverage high-order feature interactions to achieve great performance, their interaction components are  empirically predefined, which may lead to the suboptimal learning of restricted feature interactions.
As a promising approach, AFN~\cite{cheng2020adaptive} uses logarithmic neural networks (LNN)~\cite{hines1996logarithmic} to adaptively model the arbitrary-order interactions, but at the expense of restricting feature embeddings to positive real vectors, which may degrade the expressiveness of feature representations and require much more parameters to retain the performance.
Different from them, our proposed EulerNet models the feature interactions in a complex vector space by conducting the space mapping via Euler's formula.
The feature interactions in our model are adaptively learned from data without additional restrictions, which could largely improve its capacity and better balance the effectiveness and efficiency.

\paratitle{Implicit Feature Interaction Learning.}
In recent years, many deep learning based models~\cite{huang2013learning, zhang2016deep, guo2017deepfm,cheng2016wide, naumov2019deep, chen2019flen} have been proposed to model the high-order feature interactions via a deep neural network (DNN) component.
Among them, the Wide \& Deep~\cite{cheng2016wide} network combines the logit value of a linear regression model with the output of a DNN.
Besides, PNN~\cite{qu2016product} introduces an MLP to improve the output of its explicit interaction component, and NFM~\cite{he2017neural} stacks deep neural networks after FMs to model the high-order feature interactions.
Different from the explicit feature interactions, the implicit feature interactions modeled by deep neural networks lack good interpretability. 
Additionally, some recent study~\cite{rendle2020neural} has found that it is more challenging for an MLP to effectively learn the high-order feature interactions compared to using an inner product in FM.
Most deep learning based methods~\cite{guo2017deepfm,lian2018xdeepfm,pande2020field,yu2020deep} leverage the implicit feature interactions as the supplemental signal of the explicit feature interaction component.
Different from them, in EulerNet, the explicit and implicit feature interactions are learned in a unified architecture: both of them perform the linear transformations on the features in different forms (\ie the polar form for the explicit feature interactions and the rectangular form for the implicit feature interactions). 
Euler's formula establishes the relationship  between different representation forms and also builds a bridge between the explicit and implicit feature interactions.
It is observed that there exists a complementary effect between the explicit and implicit interactions in EulerNet, which enables them to promote each other and further improve the model capabilities.
\section{Conclusion}
In this paper, we proposed an adaptive feature interaction learning neural network EulerNet. 
Different from prior work, EulerNet modeled the arbitrary-order feature interactions in a complex vector space by conducting space mapping according to Euler's formula.
In EulerNet, the exponential powers of feature interactions were converted into simple linear combinations of the modulus and phase of the complex features,  enabling  it to adaptively learn the arbitrary-order feature interactions in an efficient way.
Furthermore, EulerNet integrated the implicit and explicit feature interactions into a unified architecture, which can achieve the mutual enhancement and largely boost the model capabilities.
As the major contribution, we proposed to conduct feature interaction learning in the complex vector space, which provides a way to enhance the representation capability of models and promote the feature interaction learning in this area.

As future work, we consider incorporating the user behavior features into our method, and further explore the use of attention mechanism in the complex vector space to capture more informative correlations for various recommendation tasks.



\begin{acks}
{This work was partially supported by National Natural Science Foundation of China under Grant No. 62222215 and 62102038, Beijing Natural Science Foundation under Grant No. 4222027, and  Beijing Outstanding Young Scientist Program under Grant No. BJJWZYJH012019100020098.}
\end{acks}

\bibliographystyle{ACM-Reference-Format}
\bibliography{sample-base}


\end{document}